\documentclass[12pt,english]{article}
\usepackage[english]{babel}
\usepackage{amsmath,amssymb,amscd,color}
\usepackage{amsfonts}
\usepackage{mathrsfs}
\usepackage{mathtools}
\usepackage{float}
\usepackage{graphicx}
\usepackage{url}
\usepackage{hyperref}
\usepackage{cite}
\usepackage{physics}
\usepackage{bbold}
\usepackage{braket}
\usepackage{verbatim}
\usepackage[font=footnotesize,labelfont=bf,width=.9\textwidth]{caption}
\usepackage{multirow}
\usepackage{boldline}
\usepackage{enumitem}
\usepackage{wrapfig}
\usepackage[normalem]{ulem}
\allowdisplaybreaks

\hypersetup{
    colorlinks,%
    citecolor=blue,%
    filecolor=blue,%
    linkcolor=blue,%
    urlcolor=blue
}

\makeatletter
\renewcommand\section{\@startsection {section}{1}{\z@}%
                                   {-5.5ex \@plus -1ex \@minus -.2ex}
                                   {2.3ex \@plus.2ex}%
                                   {\normalfont\large\bfseries}}
\renewcommand\subsection{\@startsection{subsection}{2}{\z@}%
                                     {-3.25ex\@plus -1ex \@minus -.2ex}%
                                     {1.5ex \@plus .2ex}%
                                     {\normalfont\bfseries}}

\numberwithin{equation}{section}

\makeatother

\newcommand{\bea}{\begin{eqnarray}}
\newcommand{\eea}{\end{eqnarray}}
\newcommand{\be}{\begin{equation}}
\newcommand{\ee}{\end{equation}}

\renewcommand{\title}[1]{\vbox{\center\LARGE{#1}}\vspace{5mm}}
\renewcommand{\author}[1]{\vbox{\center#1}\vspace{5mm}}
\newcommand{\address}[1]{\vbox{\center\footnotesize\em#1}}
\newcommand{\email}[1]{\vbox{\center\footnotesize\tt#1}\vspace{5mm}}

\begin{document}

\begin{titlepage}

 \begin{flushright}

\end{flushright}

\begin{center}

\hfill \\
\hfill \\
\vskip 1cm

\title{Near-Extremal Limits of de Sitter Black Holes}

\author{Alejandra Castro$^{a}$,  Francesca Mariani$^{b,c,d}$, and Chiara Toldo$^{b,e,f}$ 
}

\address{
${}^a$ Department of Applied Mathematics and Theoretical Physics, University of Cambridge,\\ Cambridge CB3 0WA, United Kingdom
\\
${}^{b}$
Institute for Theoretical Physics, University of Amsterdam, Science Park 904, \\
1090 GL Amsterdam, The Netherlands\\
${}^c$ Dipartimento di Fisica, Universita' degli studi di Milano–Bicocca, Piazza della Scienza 3,\\ I-20126 Milano, Italy
\\ 
${}^d$ Department of Physics and Astronomy,
Ghent University, Krijgslaan, 281-S9, 9000 Gent, Belgium
\\
${}^e$ Department of Physics, Jefferson Lab, Harvard University, 17 Oxford Street,\\ Cambridge, MA 02138, USA
\\
${}^f$ Dipartimento di Fisica, Universita' di Milano, via Celoria 6, 20133 Milano MI, Italy
}

\email{ac2553@cam.ac.uk, francesca.mariani@ugent.be, chiaratoldo@fas.harvard.edu}

\end{center}

\vfill

\abstract{We analyze the thermodynamic response near extremality of charged black holes in four-dimensional Einstein-Maxwell theory with a positive cosmological constant. The latter exhibit three different extremal limits, dubbed cold, Nariai and ultracold configurations, with near-horizon geometries AdS$_2 \times S^2$, dS$_2 \times S^2$, Mink$_2 \times S^2$, respectively. For each of these three cases we analyze small deformations away from extremality, and contrast their response. 
We also construct the effective two-dimensional theory, obtained by dimensional reduction, that captures these features and provide a more detailed analysis of the perturbations around the near-horizon geometry for each case. 
Our results for the ultracold case in particular show  an interesting interplay between the entropy variation and charge variation, realizing a different response in comparison  to the other two near-extremal limits.}

\vfill

\end{titlepage}

\eject

{
  \hypersetup{linkcolor=black}
  \tableofcontents
}
\section{Introduction}

Black holes are notorious for their semi-classical features: they are usually characterized by only a few parameters, and exhibit universal laws governing the mechanics (thermodynamics) of their horizons. In this context it is reasonable to expect an overarching principle that accounts for these laws. However, it is also known that some of these features suffer modifications depending on the surrounding of the black hole. In particular, the presence (or absence) of a cosmological constant has both quantitative and qualitative repercussions in  our understanding of black holes. Differences due to the surrounding is what we aim to explore and quantify here.

Black holes embedded in de Sitter, i.e., a gravitational theory with a positive cosmological constant, are an ideal laboratory to explore these differences. The presence of de Sitter adds a cosmological horizon which is well-known to mimic thermodynamic behaviour \cite{PhysRevD.15.2738}.\footnote{It should also be mentioned that accounting for a statistical origin of the thermodynamic properties of de Sitter is  difficult, as it has been stressed and reviewed by several authors; see for instance \cite{Witten:2001kn,Strominger:2001pn,Banks:2005bm,Anninos:2012qw,Anninos:2017eib,Coleman:2021nor}.}  It also allows for interesting generalizations, such as those explored, for instance, in \cite{Dolan:2013ft}. The presence of a cosmological horizon, plus  the existing Cauchy horizons of the black hole, provides an extra dial that will allow us to explore different thermodynamic regimes and contrast behaviour.

To be concise, we will focus on four-dimensional electrically charged black holes embedded as solutions to Einstein-Maxwell theory with a positive cosmological constant. These are the so-called Reissner-Nordstr\"om de Sitter black holes (RNdS$_4$), which have the additional property of being spherically symmetric. These configurations generally admit three horizons: an inner, outer and cosmological horizon. The confluence of these horizons, which defines an extremal limit of the original black hole, is an interesting starting point for two reasons. First, for RNdS$_4$ there are three different extremal limits \cite{Romans:1991nq,Mann:1995vb,Booth:1998gf},
dubbed {\it cold} (inner and outer horizon coincide), {\it  Nariai} (outer and cosmological horizon coincide) and {\it ultracold} (confluence of three horizons) configurations. The near-horizon geometry for each case are AdS$_2 \times S^2$, dS$_2 \times S^2$, Mink$_2 \times S^2$, respectively. Each of these instances has its own strengths and intricacies which we will highlight below; it also provides us a dial to contrast the effects of the surrounding.

Second, starting from extremality we can apply holographic tools to decode and interpret the thermodynamic responses away from extremality. This has been a powerful strategy for extremal black holes with an AdS$_2$ factor in the near-horizon geometry: deformations away from extremality define the concept of near-AdS$_2$/near-CFT$_1$ \cite{Almheiri:2014cka,Maldacena:2016upp} which leads to important insights on the quantum nature of black holes; see \cite{Mertens:2022irh} for a recent review. Here we will explore the concept of being ``near'' for the three different extremal cases of RNdS$_4$. This is where spherical symmetry comes in handy: we can build a consistent two-dimensional effective theory that captures the AdS$_2$, dS$_2$ and Mink$_2$ dynamics and its deformations. This two-dimensional theory shares many features with  Jackiw-Teitelboim (JT) gravity \cite{Jackiw:1984je,Teitelboim:1983ux}, the CGHS \cite{Callan:1992rs} and $\widehat{\rm CGHS}$ models \cite{Afshar:2019axx}, which we will review and exploit.

To summarise, we will capture and contrast various corners of black hole thermodynamics within one overarching solution, the RNdS$_4$ black hole. We will focus on semi-classical properties of the solution, and quantify them from the four-dimensional point of view. We will then provide a different perspective of these features by analyzing the holographic properties of the near-horizon geometry using the two-dimensional description. The most prominent features we find for each near-extremal configuration  are the following.
\begin{description}
    \item[Heating up cold black holes.] The AdS$_2$ factor in the near-horizon geometry controls the dynamics, which exhibits similar patterns as the analysis of near-extremal Reissner-Nordstr\"om and Reissner-Nordstr\"om in AdS$_4$ \cite{Almheiri:2016fws,Nayak:2018qej}. This is expected since the analysis is only linear, and hence mostly insensitive to the surrounding of the black hole. The thermodynamic regime is controlled by $M_{\rm gap}$, and its value has an imprint of the dS$_4$ surrounding: there is an upper bound on its value. 
    \item[Deformations of Nariai.] Here the rules are dictated by responses around dS$_2$, and we take the perspective of the static patch observer.  Our main aim in this case is to highlight how certain responses are similar and different relative to AdS$_2$: several aspects can be obtained by analytic continuation as proposed in \cite{Maldacena:2002vr,Maldacena:2019cbz}, but the interpretation and reality restrictions are delicate.
    \item[Kicking ultracold backgrounds.] This is the most novel instance of extremality, where the ultracold geometry has a natural connection to $\widehat{\rm CGHS}$ models. In this case the thermodynamic response is very different relative to the cold and Nariai cases, which we discuss in detail. Actually, it is a case where the temperature plays a minimal role at leading order. We discuss the holographic properties of our Mink$_2$ theory, where dSRN$_4$ serves as a guide in two-dimensions to dictate adequate boundary conditions and interpret the results.  
\end{description}


The paper is structured as follows. In Sec.\,\ref{sec:RN} we introduce our setup, which consists of Reissner-Nordstr\"om de Sitter black holes in Einstein-Maxwell gravity with a positive cosmological constant. We describe the space of solutions and the three different extremal limits, including their near-horizon geometries. In Sec.\,\ref{sec:JTreduction} we report the effective two-dimensional theory obtained upon reduction of Einstein-Maxwell theory on the two-sphere, and present the equations for the perturbations near-extremality  of the 2d metric and the dilaton field. In Sec.\,\ref{sec:heating_cold}, Sec.\,\ref{sec:nariai} and Sec.\,\ref{sec:ultracold} we spell out the thermodynamics near-extremality of the cold, Nariai and ultracold geometries respectively, and we compute the mass gap for these configurations. In each of these three sections we supplement our analysis with the study of the perturbations near the extremal geometry, by solving the equations of the JT-gravity like model obtained upon reduction to two dimensions. We compute then the on-shell action via holographic renormalization and we comment on the pattern of symmetry breaking and compare it among the various cases. We end with a brief summary and future directions in Sec.\,\ref{sec:conclusion}.

\section{Reissner-Nordstr\"om \texorpdfstring{dS$_4$}{dS4} black holes}\label{sec:RN}

  Reissner-Nordstr\"om black holes embedded in dS space have several interesting properties that are distinct from their counterparts in AdS or flat space. In this section we will review some of these properties based on the original work of \cite{Romans:1991nq,Mann:1995vb,Booth:1998gf}. We will focus mainly on the mechanics of its horizons, and  the accessible extremal limits.

  Our interest is in black hole solutions of Einstein-Maxwell theory with a positive cosmological constant in four-dimensions. The action is given by
\begin{equation}
I_{\rm 4D}=\frac{1}{16\pi G_N}\int \dd^{4}x \sqrt{-g}\left({\cal R}^{(4)}-2\Lambda_4 -F_{\mu\nu}F^{\mu\nu}\right)~.\label{eq:EML-action}
\end{equation}
Here $\Lambda_4=3/\ell_4^2$ is the cosmological constant, and $\ell_4^2$ is the curvature radius of $dS_4$.

An electrically charged black hole, which we will coin as RNdS$_4$, is a spherically symmetric solution of Einstein-Maxwell theory, where the line element and gauge field are
\begin{equation}
\begin{aligned}
	ds^{2}&=-V(r)\dd t^{2}+\frac{1}{V(r)}\dd r^{2}+r^{2}(\dd \theta^2 +\sin^2\theta \dd\phi^2)~,\\
A&=  \frac{Q}{r}\dd t~,  \label{dsrnds}
\end{aligned}
\end{equation}
and the blackening function in the metric is given by 
\begin{equation}
V(r)=1-\frac{2M}{r}+\frac{Q^{2}}{r^{2}}-\frac{r^2}{\ell_4^2}~.
\label{wfds1}
\end{equation}
In analogy to their flat space counterparts, we will denote the constant $M$ as the `mass' of the black hole and the constant $Q$ as the electric charge. It is also straightforward to generalize these expressions to a dyonic case, where the solution also carries magnetic charge. For simplicity, and without loss of generality, we will focus on the electric case.

The horizon structure of this black hole is dictated by the roots of $V(r)$. As it is clear from \eqref{wfds1} there are four roots, however, even when all roots are real, one of them is always located at negative $r$ and hence nonphysical (it sits behind the curvature singularity). The remaining three roots can be real and positive, which we will denote in ascending order as: $r_{-}$ is the inner horizon; $r_{+}$ is the outer horizon; $r_{c}$ is the cosmological horizon.
In this notation, we are writing \eqref{wfds1} as
\be \label{warp_factor}
V(r)=-\frac{1}{\ell_4^2 r^{2}}(r+r_++r_- +r_c)(r-r_{-})(r-r_{+})(r-r_{c})~,
\ee
where\footnote{In \eqref{constraints} we have favored $(r_\pm,\ell_4^2)$; but it is important to stress that $M$ and $Q$ are symmetric with respect to $(r_c,r_\pm)$.  One can check that $ 2\ell_4^2M=(r_++r_-)(r_++r_c)(r_-+r_c)$ and $\ell_4^2 Q^2 = r_cr_+r_-(r_c+r_++ r_-)$. }
\be
\begin{aligned} \label{constraints}
    M&=\frac{1}{2\ell_4^2}(r_++r_-)(\ell_4^2 -r_+^2-r_-^2)~,\\
    Q^2&=\frac{r_+r_-}{\ell_4^2}\left(\ell_4^2-r_+^2-r_-^2-r_-r_+\right)~,\\
    \ell_4^2&=r_c^2 +r_+^2+r_-^2+r_-r_++r_-r_c+r_cr_+~.
\end{aligned}
\ee

\begin{figure}[H]
	\centering
	\includegraphics[width=0.5\linewidth]{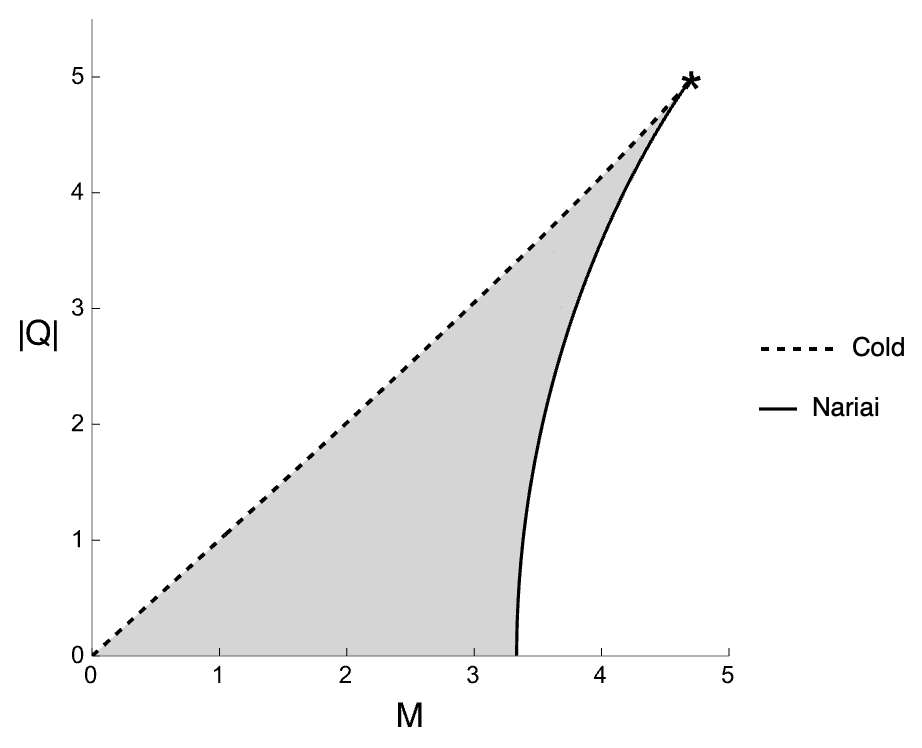}
	\caption{Shark Fin diagram for RNdS$_4$ with fixed positive value of cosmological constant $\Lambda_4=0.01$. The shaded area corresponds to black hole solutions, and the white area to naked singularities. The edges correspond to extremal black holes: the dashed line are cold solutions, the solid one are Nariai solutions. The star, where the two lines intersect, corresponds to the ultracold solution.}
	\label{SharkFin}
	\end{figure}

The space of black hole solutions is determined by the discriminant of the quartic polynomial in $V(r)$, which up to a trivial normalization reads
\begin{equation}
\begin{aligned}
    \textrm{Discr}_4&\equiv \frac{1}{16\ell_4^6} \prod_{i<j}(r_i-r_j)^2\\
    &= -16 Q^6+\ell_4^4(M^2-Q^2)+\ell_4^2(-27 M^4 + 36 M^2 Q^2 - 8 Q^4)
\end{aligned}    
\end{equation}
Here $r_i$ are all four roots in $V(r)$. Requiring   $\textrm{Discr}_4\geq 0$ and $M>0$ are  sufficient conditions to assure that three roots of the polynomial are real positive numbers. This defines for us the space of  
 solutions admitting a horizon, which we depict in Fig.\,\ref{SharkFin} for a fixed value of cosmological constant. The shaded region corresponds to a classical black hole, while the white area represents naked singularities. The shaded area is usually referred to as ``Shark Fin'' due to its shape, and we will use this nomenclature henceforth. 
	
\subsection{Black hole mechanics}\label{sec:bh-mech}

In this portion we will review mechanical properties of each physical horizon. To keep the nomenclature simple, we will refer to each quantity by its thermodynamic analog. However we do not intend to give them a statistical interpretation; do not take the analogy as an equality.

For each physical horizon $r_{\mathsf{h}}=\{r_c,r_+,r_-\}$ we can define intrinsically on them an entropy, temperature, and chemical potential, which are defined in the standard way. The area-law for the entropy is 
\be 
S_\mathsf{h}=\pi r_\mathsf{h}^2~,
\label{entropy_cosmo}
\ee 
while the Hawking temperature and electric potential read
\be
T_\mathsf{h}=\frac{1}{4\pi}|V'(r_\mathsf{h})|~, \qquad \Phi_{\mathsf{h}}=\frac{Q}{r_\mathsf{h}}~ .
\label{tcosmo}
\ee
For each horizon $r_\mathsf{h}$ we can verify that a first law is satisfied
\be 
\begin{aligned}
dM=-T_{-} dS_{-}+\Phi_{-}\, dQ~,\\
dM=\phantom{-}T_{+} dS_{+}+\Phi_{+}\, dQ~,\\
dM=-T_{c}\, dS_{c}+\Phi_{{c}}\, dQ~,
\end{aligned}
\label{1stlawbh}
\ee
where $M$ and $Q$ are the mass and  electric charge in \eqref{constraints}, and we are fixing $\Lambda_4$ ($\ell_4$) as we vary the mass, charge, and entropy. Notice that depending on the choice of $r_{\mathsf{h}}$, the first law is modified appropriately. For cosmological horizons this odd sign in \eqref{1stlawbh}  was famously observed in \cite{Gibbons:1976ue}; see \cite{Banihashemi:2022htw,Morvan:2022aon} and references within for a recent discussion.

\subsection{Three different extremal limits}\label{sec:ext}

Extremal solutions occur when two, or more, horizons coincide. Due to the presence a positive cosmological constant, it is possible to have different extremal  scenarios. For the RNdS$_4$ black hole in consideration, we have three different cases:\footnote{Historically, and in several references, the Nariai black hole refers to the case with $Q=0$ and $r_{+}=r_{c}$ \cite{1999GReGr..31..963N}. Here we define Nariai as solution with $Q\neq 0$ and $r_{+}=r_{c}$. }
\begin{description}
	\item[~~~~i. Cold black hole:] $r_{-}=r_{+}\equiv r_0$,
	\item[~~~~ii. Nariai black hole:] $r_{+}=r_{c}\equiv r_\mathsf{n}$,
	\item [~~~~iii. Ultracold black hole:] $r_{-}=r_{+}=r_{c}\equiv r_\mathsf{uc}$. 
\end{description} 
These cases describe the edges and tip of the Shark Fin in Fig.\,\ref{SharkFin}. The shared characteristic of all cases is that $T_\mathsf{h}=0$, i.e., the Hawking temperature vanishes at extremality. Each case will also have a decoupling limit, leading to an enhancement of symmetries in the near-horizon geometry. However, as we will explore below, the resulting near-horizon geometry is distinct in each case. In the following we will review the decoupling limit for each case and describe the resulting symmetries in the near-horizon region.  

\paragraph{i. Cold black hole.} The cold  solution occurs when the inner and the outer black hole horizons coincide
\begin{equation}
r_{-}=r_{+}\equiv r_0~.
\label{Coolddef}
\end{equation}
The blackening factor \eqref{warp_factor} in this case becomes 
\begin{equation}
\begin{aligned}
    V(r)_{\rm cold}=\left( 1 - \frac{r^2}{\ell_4^2} - 2 \frac{r_0 r}{\ell_4^2}   -3\frac{r_0^2}{\ell_4^2} \right)\left(1-\frac{r_{0}}{r}\right)^2~.
\end{aligned}
\label{RomansCold}
\end{equation}
Notice that between $r_0<r<r_c$ we have that $V(r)_{\rm cold}>0$.

To construct the near-horizon geometry we consider the following coordinate transformation, 
\begin{equation}
	r = r_0 +\lambda R~,\qquad t = \frac{\ell_{\rm A}^2}{\lambda} T~.
	\label{ctr}
\end{equation}
With some insight, we have introduced $\ell_{\rm A}$, which we will define below. The decoupling limit is defined as the limit where $\lambda \rightarrow 0 $, while holding $T$ and $R$ fixed; this will take us close to $r\to r_0$. Using \eqref{Coolddef} and \eqref{ctr} in \eqref{dsrnds}, the limit leads to the line element 
\begin{equation}
	ds^{2}= \ell_{\rm A}^2 \left( - R^{2} \dd T^{2}+ \frac{\dd R^{2}}{R^2}\right)+r_0^{2}\,\left(\dd\theta^2+\sin^2\theta\dd\phi^2\right)~.
    \label{dscoldext}
\end{equation}
This is the near-horizon geometry for the cold black hole: it is AdS$_2 \times S^2$. The AdS$_{2}$ radius is given by
\begin{equation}\label{ads2-cold}
	\ell_{\rm{A}}^2=\frac{r_0^{2}}{(1-6 r_0^{2}/\ell_4^2)}~,
\end{equation}
while the $S^2$ radius is the horizon radius $r_0$. Note that demanding $6r_0^2<\ell_4^2$, i.e., $\ell_{\rm A}^2>0$,  implies that $r_c>r_0$: this is consistent with our hierarchy for the roots of $V(r)$. 
It is also useful to write the charge and mass of the black hole: as a consequence of \eqref{Coolddef}, we have
\be 
Q^2_0=r_0^2\left(1-3 \frac{r_0^2}{\ell_4^2}\right)~,\qquad M_0= r_0\left(1-2\frac{  r_0^2}{\ell_4^2}\right)~.
\label{ZMRomans}
\ee
The requirement $0\leq 6r_0^2<\ell_4^2$ assures to us that $Q_0^2$ and $M_0$ are always non-negative. Hence, starting from a cold solution, we can only find a neutral solution, $Q_0=0$, by setting $r_0=0$. This is a simple way to see that the cold black hole is the dashed line in Fig.\,\ref{SharkFin}.

Finally, the field strength in this limit is also well-defined. Starting from \eqref{dsrnds} and using \eqref{ctr}, we find
\be
F=\dd A = \frac{Q_0}{r_0^2}\, \ell_{\rm A}^2\dd T \wedge \dd R~,
\ee
i.e., the field strength is proportional to the volume 2-form of AdS$_2$.

To conclude, for cold black holes we obtained an AdS$_{2}\times S^2$ factor in the metric of the near-horizon region of the extremal solution. The initial RNdS$_4$ metric had only a $u(1)$ symmetry corresponding to time translations and a $so(3)$ spherical symmetry. In the near-horizon geometry of the extremal cold solution we have an $sl(2,\mathbb{R})\times so(3)$ symmetry. Therefore the initial $u(1)$ symmetry is enhanced to a $sl(2,\mathbb{R})$ symmetry.  

 \paragraph{ii. Nariai black hole.}
 In this case the double root $r_\mathsf{n}$ refers to the coinciding horizons $r_{+}$ (outer) and $r_{c}$ (cosmological). The blackening \eqref{wfds1} factor is very similar to the cold case, where after setting $r_+=r_c=r_\mathsf{n}$, we get
\begin{equation}
\begin{aligned}
    V(r)_{\rm Nariai}=\left( 1 - \frac{r^2}{\ell_4^2} - 2 \frac{r_\mathsf{n} r}{\ell_4^2}   -3\frac{r_\mathsf{n}^2}{\ell_4^2} \right)\left(1-\frac{r_\mathsf{n}}{r}\right)^2~.
\end{aligned}
\label{RNariai}
\end{equation}
However, in contrast with \eqref{RomansCold},  for the Nariai limit the blackening factor obeys
\be\label{VN1}
V(r)_{\rm Nariai}<0~, \qquad r_-<r<r_\mathsf{n}~. 
\ee
This will be important when obtaining the resulting near-horizon geometry.

The decoupling limit is very similar to the one in \eqref{ctr}: we will consider 
\begin{equation}
	r = r_\mathsf{n} -\lambda R~,\qquad t = \frac{\ell_{\rm dS}^2}{\lambda} T~,
	\label{ctr1}
\end{equation}
and take  $\lambda\to0$ as the rest of the variables are held fixed. In comparison to \eqref{ctr}, here we have modified the sign in the radial variable: this is to demonstrate that we are reaching the horizon $r_\mathsf{n}$ from the interior of the cosmological solution (and not the inflationary patch). The parameter $\ell_{\rm dS}$ is given by
\begin{equation}\label{ds2-nar}
	\ell_{\rm{dS}}^2=\frac{r_\mathsf{n}^{2}}{(6 r_\mathsf{n}^{2}/\ell_4^2-1)}~.
\end{equation}
Notice the similarity with \eqref{ads2-cold}. However, in this case we have that $r_\mathsf{n}>r_-$, which implies that 
$6 r_\mathsf{n}^{2} >\ell_4^2$ and hence $\ell_{\rm{dS}}^2>0$. Implementing \eqref{ctr1} on the line element of RNdS$_4$, we find
\begin{equation}
ds^{2}=\ell^2_{\rm dS}\left(  R^{2} \dd T^{2}- \frac{\dd R^{2}}{R^2}\right)+r_\mathsf{n}^{2}\,\left(\dd\theta^2+\sin^2\theta\dd\phi^2\right)~.
\label{dsN}
\end{equation}
The resulting geometry is of the form dS$_2\times S^2$, where
the dS$_{2}$ radius is \eqref{ds2-nar} and the $S^2$ radius being $r_\mathsf{n}$.  One of the culprits of this change from dS$_2$, relative to AdS$_2$ in \eqref{ads2-cold}, being the signs in \eqref{VN1}. As in the cold case, the presence of a dS$_2$ factor in the near-horizon region of the extremal Nariai geometry means that we have an enhancement of symmetry with respect to the initial RNdS$_4$ metric.  
And similar to the cold case, the field strength in this limit is well-defined. Starting from \eqref{dsrnds}, we find
\be\label{Fds2}
F=\dd A = \frac{Q_\mathsf{n}}{r_\mathsf{n}^2} \,\ell_{\rm dS}^2\dd T \wedge \dd R~,
\ee
i.e., the field strength is proportional to the volume 2-form of dS$_2$. 

Finally, it is instructive to inspect the mass and the charge. Written in terms of $r_\mathsf{n}$ and $\ell_4$, we have
\be 
Q^2_\mathsf{n}=r_\mathsf{n}^2\left(1-3 \frac{r_\mathsf{n}^2}{\ell_4^2}\right)~,\qquad M_\mathsf{n}= r_\mathsf{n}\left(1-2\frac{  r_\mathsf{n}^2}{\ell_4^2}\right)~.
\label{ZMN1}
\ee
Our bounds in this case are $\ell_4^2/6< r_\mathsf{n}^{2} \leq \ell_4^2/3$. Hence the neutral solution, $Q_\mathsf{n}=0$, corresponds to $3r_\mathsf{n}^2=\ell_4^2$, for which $M_\mathsf{n}=r_\mathsf{n}/3$, as expected \cite{1999GReGr..31..963N}. 
Given this range of $r_\mathsf{n}$, the expressions in \eqref{ZMN1} lead to the solid line in Fig.\,\ref{SharkFin}.

\paragraph{iii. Ultracold black hole.}
 This case actually represents the most constrained solution related to the previous ones. The extremal ultracold solution is characterized by 
\be
r_-=r_+=r_c\equiv r_\mathsf{uc}~.
\label{UC}
\ee
Using  \eqref{UC} in \eqref{constraints}, we find
\be\label{eq:rucl}
 r_\mathsf{uc}=\frac{\ell_4}{\sqrt{6}}~,
\ee
and
\be
Q^2_\mathsf{uc}=\frac{9}{8} M_\mathsf{uc}^2=\frac{\ell_4^2}{12}~.
\ee
Hence, all scales in the black hole solution are determined by $\ell_4$. The ultracold solution is where cold black and Nariai black holes intersect, which happens when $\ell_{\rm A}=\ell_{\rm dS}\to \infty$, in accordance to \eqref{eq:rucl}. 

The blackening factor in this case reads
\be 
V(r)_{\rm ultracold}=-\frac{r^2}{6r_\mathsf{uc}^2}\left(1-\frac{r_\mathsf{uc}}{r}\right)^3\left(1+3\frac{r_\mathsf{uc}}{r}\right)~.
\ee
However, to describe the decoupling limit we need to move slightly away from this point. One way to proceed is to start from the cold case, and it is convenient to rewrite \eqref{RomansCold} as follows
\be 
V(r)_{\rm cold}=-\frac{r^2}{\ell_4^2}\left(1-\frac{r_0}{r}\right)^2\left(1-\frac{r_c}{r}\right)\left(1+\frac{2r_0+r_c}{r}\right)~,
\label{Vcold}
\ee
where we are making explicit the dependence on $r_c$, and hence the additional cosmological horizon in  $V(r)_{\rm cold}$. 
In order to capture the near-horizon region of the ultracold black hole, we start from (\ref{Vcold}) and approach the ultracold case; this means sending
\be 
r_0\rightarrow r_\mathsf{uc} -\lambda~,\qquad r_c\rightarrow r_\mathsf{uc}
+\lambda~,
\label{bro}
\ee
where $\lambda$ is the decoupling parameter. For the coordinates,  we will be performing the following transformation on the cold metric\footnote{One way to avoid going through the cold black hole, and hence avoid using \eqref{bro}, is to modify \eqref{nearhorizonuc}. We can take \eqref{UC} with $r= r_\mathsf{uc}-R_0\lambda+\sqrt{\frac{2R_0^3 }{3 r_\mathsf{uc}^{3}}}\lambda^{3/2}R$ and $t= \sqrt{\frac{3 r_\mathsf{uc}^{3}}{2R_0^3}}\frac{T}{\lambda^{3/2}}$, and this will also lead to \eqref{eq:ext-ucold-1}. Here $R_0$ is an arbitrary constant. }
\be 
r= r_\mathsf{uc}+\sqrt{\frac{2 }{3 r_\mathsf{uc}^{3}}}\lambda^{3/2}R~,\qquad t= \sqrt{\frac{3 r_\mathsf{uc}^{3}}{2}}\frac{T}{\lambda^{3/2}}~.
\label{nearhorizonuc}
\ee 
Plugging (\ref{bro}) and (\ref{nearhorizonuc}) into the metric \eqref{dsrnds}, with \eqref{Vcold}, in the limit $\lambda\to 0$ we find 
\be \label{eq:ext-ucold-1}
ds^2=-\dd T^2+\dd R^2+r_\mathsf{uc}^2 \,\left(\dd\theta^2+\sin^2\theta\dd\phi^2\right)~,
\ee
that is a geometry of the form Mink$_2\times S^2$, where the $S^2$ radius is $r_\mathsf{uc}$. This is the resulting near-horizon geometry of the ultracold black hole. The resulting field strength is 
\be\label{eq:ext-ucoldF-1}
F=\dd A = \pm\frac{\sqrt{3}}{\ell_4} \dd T \wedge \dd R~.
\ee

\section{Effective two-dimensional theory \label{sec:JTreduction}}

In the subsequent sections we will be analyzing the deviations away from the extremal limits of the RNdS$_4$ black hole, for each case described in Sec.\,\ref{sec:ext}. Since all the extremal limits correspond to geometries that are the direct product of two-manifold and round two-sphere, i.e., of the form ${\cal M}_2\times S^2$, it is convenient to construct the effective gravitational theory on  ${\cal M}_2$.

There are several references that describe the dimensional reduction of Einstein-Maxwell theory on a two-sphere; here we will be mainly following the conventions of \cite{Nayak:2018qej}---see also \cite{Larsen:2018iou,Castro:2021wzn}. The 4D action is given by \eqref{eq:EML-action}.
We will do a dimensional reduction of this theory to two-dimensions, where we take
\be\label{eq:metric4d}
\begin{aligned}
 ds^2_4&=g^{(4)}_{\mu\nu} \dd x^\mu \dd x^\nu= \frac{\Phi_0}{\Phi} g_{ab} \dd x^a \dd x^b + \Phi^2 \left(\dd\theta^2+\sin^2\theta\dd\phi^2\right)~,\\
  F&= F_{ab} \dd x^a \wedge \dd x^b ~.   
\end{aligned}
\ee
Here $g_{ab}$ is a 2D metric, and $\Phi(x)$ is a scalar field, usually coined as the {\it dilaton}; both fields depend only on the 2D coordinates $x^a$, $a,b=(0,1)$. $\Phi_0\equiv \Phi(x=x_0)$ is a constant that we use to normalize solutions and compare the 2D solutions with their four-dimensional counterparts.  Notice that we are assuming that all configurations preserve spherical symmetry, and the field strength in four dimensions, $F_{\mu\nu}$, is purely electric (i.e. supported solely by the two-dimensional metric). This is a consistent truncation of the theory, and it will suffice for our purposes.

The result of the dimensional reduction over \eqref{eq:metric4d} on \eqref{eq:EML-action} leads to the effective two-dimensional action 
\be\label{eq:2daction}
\begin{aligned}
{I_{\rm 2D}=\frac{1}{4 G_4} \int_{{\cal M}_2} \dd^2x \sqrt{-g}\Phi^2 \left( {\cal R}+ 2\frac{\Phi_0}{\Phi^3}-{2\Lambda_4}\frac{\Phi_0}{\Phi}-\frac{\Phi}{\Phi_0} F_{ab}F^{ab} \right)~.}
\end{aligned}
\ee 
Here ${\cal R}$ is the two-dimensional Ricci scalar associated to $g_{ab}$. The resulting equations of motion are as follow. The variation of the action with respect to the dilaton gives
\be\label{eq:eom1}
\begin{aligned}
{{\cal R} -  \frac{\Phi_0}{\Phi^3} -\frac{3}{2}\frac{\Phi}{\Phi_0} F^2 -\Lambda_4 \frac{\Phi_0}{\Phi} =0 ~,}
\end{aligned}
\ee
and the variation with respect to the metric leads to
\be\label{eq:eom2}
\begin{aligned}
{(\nabla_a\nabla_b -g_{ab}\square) \Phi^2 +g_{ab}\left(\frac{\Phi_0}{ \Phi} +\frac{1}{2}\frac{\Phi^3}{ \Phi_0} F^2 -\Lambda_4 {\Phi_0 \Phi} \right)=0~.}
\end{aligned}
\ee
Lastly, variation with respect to the gauge field yields:
\be \label{maxw}
\partial_a \left( \sqrt{-g} \frac{\Phi^3}{\Phi_0} F^{ab} \right) =0~.
\ee
 It is important to remark that all solutions to \eqref{eq:eom1}-\eqref{maxw} solve the equations of motion of the four-dimensional theory \eqref{eq:EML-action}.  The solution to Maxwell's equations is
\be\label{eq:F1}
F_{ab}= Q \frac{\Phi_0}{ \Phi^3}\sqrt{-g}\epsilon_{ab}~,
\ee
and $Q$ is a constant, i.e., the electric charge. 
It is also useful to record
\be
\begin{aligned}
{F_{ac}F_{b}^{~c}=Q^2  \frac{\Phi_0^2}{ \Phi^6} g_{ab}~,\qquad F^2= -2Q^2  \frac{\Phi_0^2}{ \Phi^6} ~.}
\end{aligned}
\ee
However we will not substitute this on-shell value in the 2d theory since we might not want to keep the electric charge $Q$ fixed: we will proceed with the variation of the gauge field as well.

As an example, it is constructive to write dSRN$_4$ in the language of the two-dimensional theory. Comparing \eqref{dsrnds} with \eqref{eq:metric4d}, we find that the dilaton is simply
\be \label{eq:rn1}
\Phi(x)=r~,
\ee
and 
\be\label{eq:rn2}
g_{ab}\dd x^a \dd x^b = \frac{\Phi }{\Phi_0}\left( -V(r) \dd t^2+ \frac{\dd r^2}{V(r)}\right) ~,\qquad F= \frac{Q}{r^2} \dd t\wedge \dd r~,
\ee
with $V(r)$ given in \eqref{wfds1}. It is straightforward to verify that \eqref{eq:rn1}-\eqref{eq:rn2} is a solution to \eqref{eq:eom1}-\eqref{maxw}. Notice that the electric charge of the black hole in Sec.\,\ref{sec:RN} is exactly the same as the constant entering in \eqref{eq:F1}. 

We will mostly be interested in describing the dynamics surrounding the near-horizon geometries of the three extremal cases: cold, Nariai, and ultracold. We will denote these near-horizon backgrounds as the IR solutions. From the two-dimensional perspective, IR means that we analyze solutions starting from a background with $\Phi(x)= \Phi_0$, i.e., constant dilaton background. 
When $\Phi$ is constant, we find from \eqref{eq:eom2} that
\be
Q^2 = \Phi_0^2 \left( 1-3 \frac{\Phi_0^2}{\ell_4^2} \right)~,
\ee
and hence $F_{ab}$ is a constant ($Q\Phi^{-2}_0$) times the volume element of $g_{ab}$. Equation \eqref{eq:eom1} determines the Ricci scalar to be
\be\label{eq:ads2}
{\cal R}_0 = -\frac{2}{\Phi_0^2} \left( 1- 6\frac{\Phi_0^2}{\ell_4^2}\right)~. 
\ee
That is, the metric $g_{ab}$ has constant curvature. If $6\Phi_0^2<\ell_4^2$, the solution is locally AdS$_2$,   where the curvature radius of the 2D geometry is
\be \label{radius3}
\frac{1}{\ell_{\rm A}^2}=  \frac{1}{\Phi_0^2} \left( 1- 6\frac{\Phi_0^2}{\ell_4^2}\right)~. 
\ee
In comparison to \eqref{dscoldext}-\eqref{ads2-cold}, we have $\Phi_0=r_0$ as expected.  If $6\Phi_0^2>\ell_4^2$, then the solution is locally dS$_2$, with curvature radius
\be \label{radius2}
\frac{1}{\ell_{\rm dS}^2}=  \frac{1}{\Phi_0^2} \left( 6\frac{\Phi_0^2}{\ell_4^2}-1\right) ~. 
\ee
Comparing to \eqref{ds2-nar}-\eqref{dsN}, we have  $\Phi_0=r_\mathsf{n}$. 
And if $6\Phi_0^2=\ell_4^2$, the solution is locally Mink$_2$ with $\Phi_0=r_\mathsf{uc}$, in accordance with \eqref{eq:rucl}.

In the following sections we will consider linear fluctuations around this IR background. For this purpose, we define
\be
\begin{aligned}\label{eq:lin3}
\Phi &= \Phi_0 + \lambda\, Y(x)~,\\
g_{ab}&= \bar g_{ab} + \lambda\, h_{ab}~, \\
A_{a}&= \bar A_{a} + \lambda\, \mathcal{A}_{a}~,
\end{aligned}
\ee
where $\bar g_{ab}$ is the metric for a locally AdS$_2$, dS$_2$ space or Mink$_2$ space, i.e., satisfies \eqref{eq:ads2}, and $\lambda$ is a small parameter. The fluctuations of the gauge field $\mathcal{A}_{a}$ enter via the field strength in the equations of motion. And these are determined \eqref{eq:F1}: from there we see that
 \be\label{eq:F2}
 \delta F_{ab}=  \frac{\delta Q }{\Phi_0^2}\sqrt{-\bar{g}}\epsilon_{ab} - \frac{3Q }{ \Phi_0^3} Y \sqrt{-\bar{g}}\epsilon_{ab} +  \frac{Q }{ 2\Phi_0^2} \sqrt{-\bar{g}} \epsilon_{ab} \bar{g}^{cd} h_{cd}~.
 \ee
In most of the prior literature it is common to set $\delta Q=0$, i.e., to hold the charge fixed. However, for our purpose it will be important to keep $\delta Q$ arbitrary. With this in mind, from  the equations of motion \eqref{eq:eom2} and \eqref{eq:eom1}, we find that at linear order in $\lambda$ the dynamics of $Y(x)$ and $h_{ab}$ become
\be\label{massaged1}
\begin{aligned}
{(\bar \nabla_a\bar \nabla_b -\bar g_{ab}\bar \square) Y(x)-\frac{{\cal R}_0}{ 2} \bar g_{ab} Y(x) -\frac{1}{\Phi_0^3}  \bar g_{ab} Q \delta Q =0~,}
\end{aligned}
\ee
and 
\be \label{massaged2}
\begin{aligned}
{\bar \nabla^a\bar \nabla^b h_{ab} -\bar \square h(x) - \frac{{\cal R}_0}{ 2} h(x)- \frac{12}{\Phi_0^3} \left( 1-4\frac{\Phi_0^2}{\ell_4^2} \right)Y(x) +\frac{6}{\Phi_0^4} Q \delta Q  =0~,}
\end{aligned}
\ee
where $h(x)= h_{ab}\bar g^{ab}$. In the following \eqref{massaged1}-\eqref{massaged2} will be dictating the response of the system as we move away from the IR background. And as each extremal case is discussed, we will be solving this system explicitly and connecting it to the response of the RNdS$_4$ black hole. 

\section{Heating up the cold black hole }\label{sec:heating_cold}

In this section we analyze the thermodynamic response near extremality of the first branch of solutions, the so-called cold black hole, characterized by an AdS$_2 \times S^2$ near-horizon geometry. For the cold solution the inner and outer black hole horizons coincide, $r_+ = r_- \equiv r_0$, and its conserved quantities are expressed in \eqref{ZMRomans}. 

Our analysis will encompass two perspectives, which will be contrasted. First, a perspective from the four-dimensional black hole based on the mechanics in Sec.\,\ref{sec:bh-mech}, which should be viewed as a UV derivation of the response. The second perspective will come from a two-dimensional analysis, where the analysis describes the back-reaction of the AdS$_2$ IR background Sec.\,\ref{sec:JTreduction}. We will match both derivations and discuss the holographic interpretations. 

This type of analysis has  been done  extensively in the literature for black holes whose near-horizon geometries contain an AdS$_2$ factor; see \cite{Mertens:2022irh} for a recent review. Hence our discussion here will be brief, and our aim is to set a stage to make comparisons with the Nariai and ultracold black holes.

\subsection{Near-extremality: thermodynamics and geometry}\label{sec:near-cold-thermo}

In this portion, we will quantify the response away from extremality starting from the black hole solution in four-dimensions. That is we will start from a non-extremal black hole and arrange parameters such that we are near to, but not at,  the extremal configuration.  

The elementary notion of near-extremality we will use is as a  deviation of coincident horizons. For the cold black hole this will take the form  
\begin{equation}
r_{-}=r_{0}-\lambda\,\epsilon+O(\lambda^2)~,\qquad r_{+}=r_{0}+\lambda\,\epsilon+O(\lambda^2)~,
\label{expds}
\end{equation}
where $\lambda$ is the decoupling parameter in \eqref{ctr}, and $\epsilon$ is a finite parameter. The two effects we will quantify, are the leading order response in $\lambda$ of the black hole mechanics, and how the near-horizon geometry is modified by $\lambda$ and $\epsilon$.

\paragraph{Thermodynamics.}  The natural effect of \eqref{expds} is to raise the temperature: for $r_\mathsf{h}=r_+$ we have from \eqref{tcosmo}
\be\label{eq:Tplus}
T_+ =\frac{1}{4\pi \ell_4^2 r_+^2} \left(2 r_++r_-+r_c\right)\left( r_+-r_-\right) \left(r_c-r_+\right)~,
\ee
hence the near-extremal limit raises the temperature from zero to $T_+\sim O(\lambda)$. Now, in this process we will like to keep the charge $Q$ fixed: this is not a necessary condition, but one that is consistent to take.\footnote{Throughout we will always take $\ell_4$ fixed, as done in Sec.\,\ref{sec:bh-mech}. This is also not necessary, but reasonable for the comparisons we will make among the three extremal limits. } In this case, one has to adjust the $O(\lambda^2)$ in \eqref{expds}. The result will lead to a response of $Q\sim O(\lambda^3)$ and $M\sim O(\lambda^2)$, hence making the effects of the charge sub-leading.  

Taking this into account, and using \eqref{constraints}, \eqref{eq:Tplus} and  \eqref{expds}, the response of the mass as a function of the temperature is
\begin{equation}
	M=M_{0}+\frac{T_+^{2}}{M_{\rm gap}^{\rm cold}}+O(T_+^3)~,
	\label{Mextplusgap}
\end{equation}
where the extremal mass, $M_0$, is defined in \eqref{ZMRomans}. We also identify the mass gap as 
\be \label{Mgap_cold}
M_{\rm gap}^{\rm cold}=\frac{(\ell_4^2-6r_{0}^{2})}{2\pi^{2} \, \ell_4^2 \, r_{0}^{3}} ~.  
\ee
The entropy at the outer horizon is linear in  the temperature
\begin{equation}
	S_+=\pi r_{+}^{2} 
 = S_{0}+\frac{2T_+}{M_{\rm gap}^{\rm cold}}+O(T_+^{2}) ~,
	\label{Scold1}
\end{equation}
where $S_{0}=\pi r_{0}^{2}$ is the extremal entropy. The first law  at this order is simply $dM = T_+ dS_+ + O(\lambda^3) $. 

It is worth pointing out that the change in the entropy of the cosmological horizon comes at $O(\lambda^2)$, and  it is therefore subleading with respect to the change in entropy at the outer horizon. For this reason, we can consider this an ensemble of fixed charge and fixed cosmological horizon area.

\paragraph{Near-horizon geometry.}
As anticipated, separating the inner and the outer black hole horizons by a small amount increases the temperature, and hence, we see an increase in the entropy. There is as well an effect in the near-horizon geometry, which we now quantify. 

To see the form of the near-horizon geometry of this configuration we take into consideration the displacement of the horizons in \eqref{expds} as we take the decoupling limit. Adapting slightly \eqref{ctr},\footnote{This adjustment is just for aesthetic reasons, i.e., to preserve the form $g_{RR}=\ell_{\rm A}^2/R^2$ as we take the limit $\lambda\to 0$.} we now have
\be\label{ctr:near}
r= r_0 +\lambda \left(R+ \frac{\epsilon^2}{4} R^{-1}\right) ~, \qquad t=\frac{\ell_{\rm A}^2}{\lambda}T~.
\ee
Using \eqref{expds} and \eqref{ctr:near} on \eqref{dsrnds}, and taking $\lambda\to 0$, we find
\begin{equation}\label{eq:near-cold}
\begin{aligned}
	ds^{2}&=\ell_{\rm A}^2 \left( -R^{2}\left(1-\frac{\epsilon^{2}}{4R^2}\right)^2\dd T^{2}+\frac{\dd R^{2}}{R^{2}} \right)+r_{0}^{2}
\,\left(\dd\theta^2+\sin^2\theta\dd\phi^2\right)~,\\
 F &=\frac{Q_0}{r_0}\, \ell_{\rm A}^2\left(1-\frac{\epsilon^{2}}{4R^2}\right) \dd T\wedge \dd R~,
\end{aligned}
\end{equation}
where $\ell_A$ is defined in (\ref{ads2-cold}).
This is an instance of a \textit{nearly}-$AdS_2$ geometry, which arises as the near-horizon region of the near extremal solution. This geometry is locally AdS$_2$, which globally breaks some of the isometries. These are restored if we take the extremal limit, corresponding to $\epsilon\rightarrow 0$. 

It is also important to quantify the leading order response, and compare it with the holographic properties we will discuss shortly.  Using the two-dimensional notation, we will parameterize the first response in $\lambda$ as  
	\begin{equation}
	\begin{aligned} \label{ads2pluspertCE}
	ds^2&= \left(\bar{g}_{ab} + \lambda\,  h_{ab}\right)  \dd x^a \dd x^b+ \left(\Phi_0^2 + 2\lambda \Phi_0 \, Y \right)\left(\dd\theta^2+\sin^2\theta\dd\phi^2 \right)+\cdots ~,\\
 A &= \bar A_{a} \dd x^a + \lambda\, {\cal A}_a \dd x^a + \cdots~,
	\end{aligned}
	\end{equation}
that is, there is a response from the AdS$_2$ background ($h_{ab}$, ${\cal A}_a$) and the size of the two-sphere (${Y}$). Here  $\bar{g}_{ab}$ is the locally AdS$_2$ background, and $\bar A_a$ is the gauge field associated to $F$ in \eqref{eq:near-cold}.  For the near-extremal cold solution we find
\be\label{eq:Ycold}
\Phi_0 = r_0~, \qquad Y(x)=R+ \frac{\epsilon^2}{4} R^{-1} ~.
\ee
Note that we are not explicitly reporting on the profile of $ h_{\mu\nu}$, although it is straightforward to extract. This is because  $h_{\mu\nu}$ is not an independent degree of freedom. As it will be evident in Sec.\,\ref{sec:hol-cold}, its profile is determined by a choice of gauge and the dynamics of $Y$. Since we are holding the charge $Q$ fixed, the response of ${\cal A}_a$ is also dictated by $ Y$ and $h_{ab}$; see \eqref{eq:F2}.

\subsection{Holographic analysis}\label{sec:hol-cold}

In this portion we will analyze the extremal cold solutions, and their response near-extremality, from the two-dimensional perspective. We will first report the solutions for the linearized perturbations and then we analyze the renormalized action.
This follows closely the analogous derivations in \cite{Cvetic:2016eiv,Castro:2018ffi,Castro:2019vog}, which we refer to for more details.

We will cast our solutions in a radial gauge, where we take 
\be\label{eq:gauge-cold}
ds^2= \dd\rho^2 + \gamma_{TT} \dd T^2~, \qquad A=A_T \,\dd T~.
\ee 
For a cold black hole, the appropriate IR background
is \eqref{radius3}: the locally AdS$_2$ solution.  For this case we will cast the metric as
\be \label{bg_cold}
\overline{g}_{ab} \dd x^a \dd x^b= \dd\rho^2 + \bar{\gamma}_{TT} \dd T^2\,, \qquad \bar \gamma_{TT} = - \left(\alpha(T) e^{\rho/\ell_{\rm A}} + \beta(T) e^{-\rho/\ell_{\rm A}}\right)^2\,,
\ee
and the background gauge field reads
\be\label{form_gauge}
\bar A=\bar A_{T}\dd T = \mu(T)\dd T -\frac{Q \ell_{\rm A}}{\Phi_0^2} \left(\alpha(T) e^{\rho/\ell_{\rm A}} - \beta(T) e^{-\rho/\ell_{\rm A}}\right)\dd T~.
\ee
This solution is locally AdS$_2$ for arbitrary metric functions $\alpha(T)$ and $\beta(T)$; note that $\mu(T)$ is a pure gauge function. In comparison to \eqref{eq:near-cold} we have
\be\label{eq:compare12}
R= e^{\rho/\ell_{\rm A}}~,\qquad \alpha(T)_{\rm cold}= \ell_{\rm A} ~, \quad \beta(T)_{\rm cold}= -\ell_{\rm A} \frac{\epsilon^2}{4}~.
\ee

The response for this background will be parameterized by $Y(x)$ and $h_{ab}$ as defined in \eqref{eq:lin3}; since we are taking $\delta Q=0$, the response of the gauge field is fixed by \eqref{eq:F2}. The profile of the perturbations comes from solving the linearized equations given in \eqref{massaged1}-\eqref{massaged2}.  Starting with the solution to \eqref{massaged1}, in our conventions the dilaton reads
\be
Y(x)= \nu(T)e^{\rho/\ell_{\rm A}} + \theta(T) e^{-\rho/\ell_{\rm A}}~,
\ee
with
\begin{equation}
\begin{aligned}\label{eq:beta-nu}
    \beta(T) &= -\frac{\ell_{\rm A}^2}{4} \frac{\alpha}{ \partial_T \nu} \partial_T \left( \frac{1}{\nu} \left( c_0 + \frac{(\partial_T \nu)^2}{\alpha^2} \right) \right) ~, \\
\theta(T) & =  -\frac{\ell_{\rm A}^2}{4 \nu} \left(c_0 + \frac{(\partial_T \nu)^2}{\alpha^2} \right)~,
\end{aligned}
\end{equation}
where $c_0$ is a constant. Here we have chosen to express the subleading components $\beta(T), \theta(T)$ of the background metric and fluctuations in terms of the leading source terms $\alpha(T), \nu(T)$. Comparing with \eqref{eq:Ycold}, we have
\be\label{eq:compare34}
\nu(T)_{\rm cold}=1 ~,\qquad \theta(T)_{\rm cold}=\frac{\epsilon^2}{4}~.
\ee

Finally we report on the metric perturbations. Once we plug in the solution for the dilaton, the equation for the metric \eqref{massaged2} is straightforwardly solved by
\be\label{eq:g1}
h_{TT} =
-4\frac{\ell_{\rm A}^2}{ \Phi_0^3} \left(1-4\frac{\Phi_0^2}{\ell_4^2}\right) \left(  \bar\gamma_{TT} \, Y(x) -2 \ell_{\rm A}^2 \sqrt{-\bar \gamma}\, {\partial_T} \left( \frac{\partial_T \nu(T) }{\alpha (T)} \right) \right) \,.
\ee
Notice that we have focused on the inhomogeneous part of the solution, given that the homogeneous part can be absorbed in the arbitrary functions $\alpha(T)$ and $\beta(T)$ appearing in the background metric solution \eqref{bg_cold}. 

To sum up, the solution to the system \eqref{massaged1}-\eqref{massaged2} are given in function of two functions alone, $\alpha(T)$ and $\nu(T)$, appearing as (finite) source in the background metric and (infinitesimal) source for the irrelevant operator dual to the dilaton $Y$, whose conformal dimension is $\Delta=2$. At the linear level, the analysis of near-AdS$_2$ is qualitatively insensitive to the embedding inside of this theory inside of dS$_4$. That is, we do not see any significant effect of the four dimensional cosmological constant in this sector.

\subsubsection{Renormalized action}

In order to compute the renormalized on-shell action we consider the effective 2D action \eqref{eq:2daction}, 
and we plug in the solution constructed between \eqref{eq:gauge-cold}-\eqref{eq:g1}. The range of integration is taken to be a finite radial value $\rho_h$ in the IR and a cutoff regulator $\rho_0$ in the UV of AdS$_2$. The regulated action is divergent as we send $\rho_0$ to infinity, therefore it needs to be supplemented by additional boundary counterterms, which guarantee a
Dirichlet boundary problem (the Gibbons-Hawking-York term $I_{\rm GH}$) and to remove residual divergencies ($I_{\rm ct}$) once the cutoff is removed:\footnote{Recall that we are setting $G_N=1$.}
\begin{equation} \label{ct_cold_r}
    I_{\rm ct} = - \frac{1}{ \, \ell_{\rm A}} \int \dd T \,  \sqrt{-\gamma}\,\Phi^2~,  \qquad   I_{\rm GH} = \frac{1}{2 } \int \dd T \,  \sqrt{-\gamma}\, \Phi^2\,K ~.
\end{equation}

These counterterms are appropriate to a 2-dimensional spacetime ${\cal M}_2$ with a one-dimensional boundary $\partial {\cal M}_2$ with induced metric $\gamma_{ab}$. The Gibbons-Hawking surface term is given in terms of the trace of the extrinsic curvature $K_{ab}$ of the boundary, $K_{ab} = -\frac12 (\nabla_{a} n_{b} + \nabla_{b} n_{a})$, 
where $n^{a}$ is the outward-pointing vector normal on $\partial {\cal M}_2$. 
Moreover, given the form of the gauge field \eqref{form_gauge}, one can see that the leading component of $A_T$ is not the source term $\mu(T)$, but  the term proportional to the volume of AdS$_2$. In order to have a well defined variational principle in terms of the source we have to perform a double Legendre transform \cite{Cvetic:2016eiv} which has the aim of both cancelling the volume term in the conjugate momenta to the gauge field, and impose Dirichlet boundary conditions. We collect below the final form for the action, but more details can be found in \cite{Cvetic:2016eiv,Castro:2018ffi,Castro:2019vog}. 

In the final form for the renormalized on shell action the dependence on the regulator $\rho_0$ drops out and the result is finite and depends on the source functions $ \alpha(T), \nu(T), \mu(T)$ appearing explicitly in the solution. Its value is
\be
I_{\rm on-shell-cold} = -\ell_{\rm A}  \Phi_0 \, \lambda \int \dd T \left(\frac{ 4 c_0 \alpha (T)}{ \nu (T)}+\frac{ \nu '(T)^2}{\alpha (T) \nu (T)}  - \mu(T) \frac{Q}{\Phi_0^3}  \right) + I_{\rm global} \,,
\ee
where $I_{\rm global}$ denotes the value of the integral evaluated at the horizon $\rho_h$, whose explicit form is not necessary for our purposes.

Following the reasoning in \cite{Castro:2018ffi}, we can see that the renormalized action is invariant under infinitesimal time reparameterizations and $U(1)$ gauge transformations. The functions $ \alpha(T)$, $\nu(T)$, and  $\mu(T)$ are pure gauge and can be traded for the three independent functions that generate residual gauge symmetries, but the on-shell action actually depends only on one of them, which we call $\sigma(T)$, which generates a boundary Weyl transformation. Without loss of generality, therefore, we can parameterize the sources as a finite Weyl transformation starting from the reference point with $\alpha=1$, $\nu=1$, $\mu=\mu_0$, namely
\begin{equation}\label{eq:weyl}
    \alpha(T) = e^{\sigma(T)/\ell_{\rm A}}, \qquad \nu(T) = e^{\sigma(T)/\ell_{\rm A}},  \qquad \mu(T) = \mu_0\,.
\end{equation}
Inserting these in the on-shell action, the latter boils down to this simple expression:
\begin{equation} \label{onshell_final_def}
    I_{\rm on-shell-cold} =   \ell_{\rm A}  \Phi_0 \, \lambda \int \dd T \left( -{4 c_0} + 2  \{ \tau(T),T \} + \frac{Q \mu_0}{\Phi_0^3} \right) + I_{\rm global}~.
\end{equation}
 To arrive at expression \eqref{onshell_final_def}, we have parameterized $\sigma(T)$ in terms of an arbitrary auxiliary function $\tau(T)$ as
\begin{equation}
\sigma(T) = \ell_{\rm A} \log \partial_T \tau(T)~,
\end{equation}
and added a total derivative term. In the integral we have introduced 
\begin{equation}
    \{ \tau(T),T \} \equiv \frac{\partial_T^3 \tau}{\partial_T \tau} - \frac32 \left(\frac{\partial_T^2 \tau}{\partial_T \tau} \right)^2~,
\end{equation}
i.e, the Schwarzian derivative. 
Unsurprisingly, the response of the system under Weyl transformations of the boundary metric manifests at the level of the on-shell action in the appearance of the Schwarzian derivative.

We are ready now to briefly make contact with the thermodynamic analysis due to the linear response induced by $Y(x)$. We will be working with \eqref{eq:compare12} and \eqref{eq:compare34},\footnote{In order to cast the metric \eqref{dscoldext} in the form of the background $\overline{g}_{ab}$ with \eqref{bg_cold}, we need to re-scale the time coordinate in \eqref{ctr} by $T \rightarrow T/\ell_{\rm A}$, effectively resulting in the multiplicative $\ell_{\rm A}$ factor in the on-shell action.} for which the background metric is \eqref{bg_cold}.  This is a solution that contains a horizon at 
\be
\bar\gamma_{TT}(\rho = \rho_h) =0 \qquad \rightarrow \qquad e^{2\rho_h/\ell_{\rm A}} = -\frac{\beta_{\rm cold}}{\alpha_{\rm cold}}\,.
\ee
The associated temperature in 2D is defined as
\be \label{T2}
T_{2D} = \frac{1}{2\pi} \partial_{\rho} \sqrt{-\gamma}|_{\rho_h} =  \frac{\sqrt{-\alpha_{\rm cold}\beta_{\rm cold}}}{\pi} = \frac{\epsilon}{2\pi}~.
\ee
Notice that its relation to \eqref{eq:Tplus}, near-extremality, is $T_+= \frac{\lambda}{\ell_{\rm A}^2}T_{2D}$ (in accordance to the change of time coordinate in \eqref{ctr:near}). The entropy is found by evaluating the dilaton at the horizon: 
\be
\begin{aligned}
S_{2D} &= \pi \Phi(x)^2_{\rm horizon}\\
     &=  \pi \Phi_0^2 + 2\pi \Phi_0 \lambda Y(x)_{\rm horizon}  + \cdots\\
     &= \pi \Phi_0^2 + 4\pi^2 \Phi_0 \ell_{\rm A}^2 T_+ + \cdots
\end{aligned}
\ee
After using the values reported here, it is straightforward to check that this agrees with \eqref{Scold1} and \eqref{Mgap_cold}, where in the 2D language we have 
\be
M_{\rm gap}^{\rm cold}=\frac{(\ell_4^2-6r_{0}^{2})}{2\pi^{2} \, \ell_4^2 \, r_{0}^{3}} = \frac{1}{2\pi^{2} \ell_{\rm A}^2\Phi_0}~.
\ee
As shown in \cite{Maldacena:2016upp}, this linear response in temperature arises from the Schwarzian effective action in \eqref{onshell_final_def}, where $(M_{\rm gap}^{\rm cold})^{-1}$ is proportional to the coupling in front of the action.  Although all derivations here follow closely the RN black hole in Mink$_4$ and AdS$_4$, one small new aspect here is that $M_{\rm gap}^{\rm cold}$ has a maximum value, and when  $\ell_4^2=6r_{0}^{2}$, which is the ultracold limit, the mass gap vanishes. That is, $M_{\rm gap}^{\rm cold}$ is a decreasing function of the charge $Q$. 

\section{Deviations away from extremality for Nariai }\label{sec:nariai}

In this section we analyze the response away from extremality for the Nariai black hole. This was the solution with $r_+=r_c\equiv r_\mathsf{n}$, i.e., the outer and cosmological horizons coincide. The properties of the extremal solution are described in \eqref{RNariai}-\eqref{ZMN1}. A key feature here is that the near-horizon geometry for the Nariai solution is dS$_2\times S^2$, where the de Sitter radius is
\be
\ell^2_{\rm dS}=\frac{r_{\mathsf{n}}^2 \ell_4^2}{6 r_{\mathsf{n}}^2 - \ell_4^2}~,
\ee
and the key restriction to recall in this case is 
\be \label{cnd_nar}
6 r_{\mathsf{n}}^2 > \ell_4^2~.
\ee
This will be important as we contrast the cold black hole relative to Nariai: many aspects are shared by dS$_2$ and AdS$_2$, but small differences are key.

There are several recent analyses of dS$_2$, and near-dS$_2$, that apply to the Nariai limit of Schwarzschild dS$_4$ black holes, and studies from the perspective of a two-dimensional theory with a running dilaton; see for example \cite{Maldacena:2019cbz,Moitra:2022glw,Svesko:2022txo,Anninos:2022hqo}. Our presentation here will be brief---and more details are covered in the references---since the main purpose for us is to contrast this scenario with the responses in the cold and ultracold cases.   

\subsection{Near-extremality: thermodynamics and geometry}\label{sec:near-nariai-thermo}

\paragraph{Thermodynamics.} In analogy to Sec.\,\ref{sec:near-cold-thermo}, to go slightly beyond extremality, we displace $r_{+}$ and $r_{c}$ around $r_{\mathsf{n}}$ as follows
\begin{equation}
		r_{+}=r_{\mathsf{n}}-\lambda\epsilon+O(\lambda^{2})~,\qquad r_{c}=r_{\mathsf{n}}+\lambda\epsilon+O(\lambda^{2})~.
  \label{expdsnariai}
\end{equation} 
As expected, this deviation ignites a balanced temperature at outer horizon $(T_+)$ and the cosmological  horizon $(T_c)$, with both of them linear in $\lambda$ at leading order. That is,  $T_+=T_c\sim O(\lambda)$, only at leading order.

To follow the parallel with Sec.\,\ref{sec:near-cold-thermo}, here we can also fix the charge $Q$ of the black hole, which sets constraints on the higher order terms in \eqref{expdsnariai}. With this choice, and taking the perspective of the cosmological horizon, we find the following mechanical response:
\begin{equation}\label{eq:MS12}
	M=M_{\mathsf{n}}+\frac{T_c^{2}}{M_{\rm gap}^{\rm n}}+\cdots~,
\qquad 
	S_c=S_{\mathsf{n}}-\frac{2T_c}{M_{\rm gap}^{\rm n}}+\cdots~,
\end{equation}
where $M_{\mathsf{n}}$ is given in \eqref{ZMN1}, $S_{\mathsf{n}}=\pi r_\mathsf{n}^2$, and 
 \be\label{eq:gapn}
M_{\rm gap}^{\rm n}=\frac{(\ell_4^2-6r_{\mathsf{n}}^{2})}{2\pi^{2} \, \ell_4^2 \, r_{\mathsf{n}}^{3}} =-\frac{1}{2\pi^{2}\,\ell_{\rm dS}^2\, r_{\mathsf{n}}}~<0~.
\ee
Very crucially here the mass gap is negative! And one should expect this: for fixed $Q$, starting from the extremal Nariai solution represented by the right edge of the diagram in Fig. (\ref{SharkFin}), we can only decrease the mass if we want to find physical solutions.

We can also take the perspective of the outer horizon, which gives
\begin{equation}\label{eq:MS123}
	M=M_{\mathsf{n}}+\frac{T_+^{2}}{M_{\rm gap}^{\rm n}}+\cdots~,
\qquad 
	S_+=S_{\mathsf{n}}+\frac{2T_+}{M_{\rm gap}^{\rm n}}+\cdots~,
\end{equation}
and the same value of mass gap in \eqref{eq:gapn}. In this case both the entropy and mass decrease! As discussed in 
\cite{Svesko:2022txo,Anninos:2022hqo}, this can be viewed as an instability of the outer horizon, while the cosmological horizon is stable to the near-extremal perturbation.\footnote{In the cold case a similar effect to \eqref{eq:MS123} occurs at the inner horizon. It is usually not interesting to  highlight it since one tends to not place an observer between $r_-$ and $r_+$, and it is known that the inner horizon is unstable. For a cosmological horizon it is relevant to discuss the perspective of the static patch observer for whom both the outer and cosmological horizon are present.}

It is interesting to compare this thermodynamic response to those proposed originally in \cite{Bousso:1996au}, and more recently used in \cite{Morvan:2022aon}. The idea is to use a different normalization of time-like Killing vector to define the potentials. In particular, one would modify \eqref{tcosmo} to 
\be
T_{\mathsf{h},{\rm mod}}=\frac{1}{4\pi \sqrt{V(r_{\cal O})}}|V'(r_\mathsf{h})|~, 
\qquad \Phi_{\mathsf{h},{\rm mod}}=\frac{1}{\sqrt{V(r_{\cal O})}}\frac{Q}{r_\mathsf{h}}~ .
\label{tcosmo-mod}
\ee
Here $r_{\cal O}$ is defined as a point where $V'(r_{\cal O})=0$. With these definitions of potentials the first law as stated in \eqref{1stlawbh} does not hold; instead one has 
\be 
\begin{aligned}
T_{+,{\rm mod}} dS_{+}+T_{c,{\rm mod}}\, dS_{c}+\Phi_{+,{\rm mod}}\, dQ -\Phi_{{c},{\rm mod}}\, dQ=0~,
\end{aligned}
\label{1stlawbh-mod}
\ee
which relates the temperatures and potential of both horizons. 
At extremality, one has that $r_+=r_c=r_{\cal O}=r_{\mathsf{n}}$, and one of the appeals of \eqref{tcosmo-mod} is that the temperature remains finite. In particular, one finds
\be\label{eq:dS2-temp}
T_{\mathsf{n}}\equiv T_{\mathsf{h},{\rm mod}} \big|_{r_+=r_c=r_{\cal O}=r_{\mathsf{n}}}=\frac{1}{2\pi \ell_{\rm dS}}~.
\ee
Recall that the near-horizon geometry is of the form dS$_2 \times $S$^2$, with radius \eqref{ds2-nar}: indeed \eqref{eq:dS2-temp} is the temperature associated to free falling observers in dS$_2$.
However, we notice that with the definition \eqref{tcosmo-mod} the electric potential $\Phi_{\mathsf{h},{\rm mod}}$ is not finite in the extremal limit, it diverges. Thereby this definition is not suitable for our analysis of the thermodynamics, and deviations away from extremality. 

\paragraph{Near-horizon geometry.}
The near-horizon region is reached performing the usual coordinate transformation	\eqref{ctr1} combined with \eqref{expdsnariai}, where we will make just a small modification 
\begin{equation} \label{expnariainext}
	r = r_\mathsf{n} \pm \lambda R~,\qquad t = \frac{\ell_{\rm dS}^2}{\lambda} T~.
\end{equation}
In contrast to \eqref{ctr:near}, we have not modified the radial diffeomorphism since we want to reflect the static patch below. We have added a ``$\pm$'' to illustrate the difference of the displacement relative to the outer or cosmological horizon. Replacing (\ref{expnariainext}) and \eqref{expdsnariai} in (\ref{dsrnds}), and taking the decoupling limit $\lambda\rightarrow0$ while holding $T$, $R$ and the sphere fixed, we find
\begin{equation}\label{eq:near-ext-N}
	ds^{2}= \ell_{\rm dS}^2 \left( (R^{2}-\epsilon^{2}) \dd T^{2}-\frac{\dd R^{2}}{(R^{2}-\epsilon^{2})} \right)+r_{\mathsf{n}}^{2}\,\left(\dd\theta^2+\sin^2\theta\dd\phi^2\right)~,
\end{equation} 
and the field strength is given by \eqref{Fds2}. The near-horizon geometry of the near-extremal Nariai solution contains a \textit{nearly}-dS$_2$ factor. Taking the extremal limit $\epsilon\rightarrow 0$ we indeed restore the dS$_2$ factor we had in the extremal case \eqref{dsN}. 

Notice that when obtaining the line element \eqref{eq:near-ext-N}, a reasonable choice is to expand the blackening factor $V(r)$ for $r_+<r<r_c$. This restricts $-\epsilon<R<\epsilon$, and hence the result of the decoupling limit is the static patch of dS$_2$ where the Euclidean geometry is locally a sphere. One could
consider instead $r>r_c$, i.e., to have an observer on the inflationary patch, and then we would have $R>\epsilon$.

Next, let us consider the static patch of  \eqref{eq:near-ext-N} in Euclidean signature. A simple change of coordinates gives
\begin{equation}\label{eq:near-ext-N-Euc}
	ds^{2}= \ell_{\rm dS}^2 \left( (1-\tilde R^{2}) \dd t_E^{2}+\frac{\dd \tilde R^{2}}{(1-\tilde R^{2})} \right)+r_{\mathsf{n}}^{2}\,\left(\dd\theta^2+\sin^2\theta\dd\phi^2\right)~,
\end{equation}
where $R= \epsilon \tilde R$ and $T=i\epsilon^{-1} t_E$. That is we have reabsorbed the near-extremal parameter $\epsilon$ in the coordinates.  Now the space is explicitly the direct product of two $S^2$'s, one of radius $\ell_{\rm dS}$ and the other $r_{\mathsf{n}}$. The temperature $T_{\mathsf{n}}$ in \eqref{eq:dS2-temp} is then the periodicity of $t_E$ that is appropriate for \eqref{eq:near-ext-N-Euc}. However, it is important to recognise that this temperature came about from having a finite value of $\epsilon$. With this, it is natural to argue that the temperature of dS$_2$  should tie together the thermodynamic response in  \eqref{eq:MS12}, and hence this is how the modified notions of temperature in \eqref{tcosmo-mod} should be interpreted.\footnote{Notice that in the near-extremal limit, $T_{\mathsf{c}}=\lambda \epsilon \ell_{\rm dS}^{-2} T_{\mathsf{n}}$ which is compatible with the changes of time coordinates in the decoupling \eqref{expnariainext}. That is, the modified definitions in \eqref{tcosmo-mod} simply correspond to the IR observer in the throat.} This parallels the analysis done for the cold case in Sec. \ref{sec:near-cold-thermo}.

As we did in \eqref{ads2pluspertCE}, we also report on the leading order response. Using the two-dimensional notation, we will parameterize the first response in $\lambda$ as  
	\begin{equation}
	\begin{aligned} \label{ds2pluspertCE}
	ds^2&= \left(\bar{g}_{ab} + \lambda\,  h_{ab}\right)  \dd x^a \dd x^b+ \left(\Phi_0^2 + 2\lambda \Phi_0 \, Y \right)\left(\dd\theta^2+\sin^2\theta\dd\phi^2 \right)+\cdots ~,\\
 A &= \bar A_{a} \dd x^a + \lambda\, {\cal A}_a \dd x^a + \cdots~,
	\end{aligned}
	\end{equation}
that is, there is a response from the dS$_2$ background ($h_{ab}$, ${\cal A}_a$) and the size of the two-sphere (${Y}$). Here  $\bar{g}_{ab}$ is the locally dS$_2$ background, and $\bar A_a$ is the gauge field associated to $F$ in \eqref{Fds2}.  For the near-extremal Nariai solution we find
\be\label{eq:Yn}
\Phi_0 = r_\mathsf{n}~, \qquad Y(x)=\pm R~.
\ee
Here the choice of positive or negative sign would be important in determining which horizon is being deformed: the plus sign will lead to the mechanics in \eqref{eq:MS12} and the minus sign to \eqref{eq:MS123}.

\subsection{Two-dimensional analysis} \label{sec_dS}

In this last portion we will discuss the solution to the linear equations \eqref{massaged1}-\eqref{massaged2}. We will adopt a notation very similar to Sec.\,\ref{sec:hol-cold}, so the contrast with the counterpart of near-AdS$_2$ is manifest. We take the following value for the background 2d metric
\be\label{dSinfl}
ds_2^2 = -\dd\rho^2 + \bar\gamma_{TT}\, \dd T^2~, \qquad \bar \gamma_{TT} = \left(\alpha(T) e^{\rho/\ell_\mathsf{dS}} + \beta(T) e^{-\rho/\ell_{\mathsf{dS}}}\right)^2~,
\ee
which differs from formula \eqref{bg_cold} only by an overall sign, and the presence of $\ell_{\mathsf{dS}}$ instead of $\ell_{\rm A}$. Hence, here $\rho$ is time and $T$ is a spatial direction. The metric \eqref{dSinfl} can be regarded as a generalization of global coordinates for dS$_2$. For the background gauge field we have
\be
\bar A_{T} = \mu(T) -\frac{Q \ell_{\mathsf{dS}}}{\Phi_0^2} \left(\alpha(T) e^{\rho/\ell_{\mathsf{dS}}} - \beta(T) e^{-\rho/\ell_{\mathsf{dS}}}\right)~.
\ee
Notice that the solutions for Nariai, both the background and perturbations, can be easily found by noticing that the configuration is formally equivalent to the cold one, upon performing the transformation $\rho \rightarrow i \rho$ and $\ell_{\mathsf{dS}} \rightarrow i \ell_{\mathsf{A}}$. This takes Lorentzian dS$_2$ to Euclidean AdS$_2$. However, the important subtleties come from imposing reality conditions on the various arbitrary functions that appear as we solve the system.   

By adopting the same procedure as in Sec.\,\ref{sec:hol-cold}, we start by analyzing the solution to \eqref{massaged1} for $\delta Q=0$. The solution for the dilaton reads
\be
Y(x) = \nu(T) e^{\rho/\ell_{\mathsf{dS}}} + \theta(T) e^{-\rho/\ell_{\mathsf{dS}}}~,
\ee
with
\be
\beta(T) = \frac{\alpha (T) \theta '(T)}{\nu '(T)}~, \qquad  \theta = \frac{c_n}{\nu(T)}-\frac{\ell_{\mathsf{dS}}^2 \left(\nu'(T)\right){}^2}{4 \alpha (T)^2 \nu(T)}~,
\ee
and $c_n$ a constant. The metric perturbation is
\be
\sqrt{-\gamma_1} = \frac{4 \ell_{\mathsf{dS}^2} \left(4 Q^2-\Phi_0^2\right)}{3 \Phi_0^5} \left( \sqrt{-\bar \gamma} \, Y(x) +2 \ell_{\mathsf{dS}}^2 {\partial_T } \left( \frac{ \nu'(T) }{\alpha (T)} \right) \right)~.
\ee

We have displayed the solutions in a coordinate system adequate for the inflationary patch of dS$_2$. However, the responses are also interesting from the static patch perspective, as reflected in our discussion in Sec.\,\ref{sec:near-nariai-thermo}.  To move to the static patch we first need to extend $\rho$: this can be done by the coordinate change  
\be\label{eq:coshR}
\cosh \frac{\rho}{\ell_{\mathsf{dS}}}  = \frac{R}{\ell_{\mathsf{dS}}}~, 
\ee
and select
\be
\alpha_{\rm static} = -\beta_{\rm static}=\frac{\ell_{\rm dS}}{2}~, 
\ee
where now \eqref{dSinfl} becomes
\be \label{stat_p}
ds^2 = -\ell_{\mathsf{dS}}^2\left( 1-\cfrac{R^2}{\ell_{\mathsf{dS}}^2} \right) \dd T^2 + \cfrac{\dd R^2}{\left(1-\cfrac{R^2}{\ell_{\mathsf{dS}}^2}\right)} ~.
\ee
It is important to emphasize that here $R\geq \ell_{\mathsf{dS}} $. The solution for the dilaton in this case is linear in the radial coordinate,
\be
Y(x) =  R~.
\ee
This is the same solution previously found in \cite{Moitra:2022glw,Svesko:2022txo}, and shows that a metric of the form \eqref{stat_p} can be obtained via a suitable near-extremal limit starting from a Nariai configuration.

The delicate aspect of this construction is to extend the coordinates to cover an observer that is \emph{inside} the cosmological horizon; this requires starting with generic complex functions $\alpha(T)$ and $\beta(T)$, and then imposing non-trivial reality conditions on the metric on the static patch.\footnote{In more detail, one way to reach the static patch based on the present analysis is the following. Consider applying  \eqref{eq:coshR} when $R< \ell_{\mathsf{dS}}$; in this case $\rho/\ell_{\mathsf{dS}}$ is complex. Imposing  $\beta(T)=\alpha^*(T)$ keeps the line element real.}  Alternatively, we could adapt \eqref{dSinfl} to coordinates that accommodate this portion of the geometry. For example, one would start from    
\be\label{dS-static-1}
ds_2^2 =\cfrac{\dd R^2}{\left(1-\cfrac{R^2}{\ell_{\mathsf{dS}}^2}\right) } -\ell_{\mathsf{dS}}^2\left(   \alpha_{\rm s}(T)\cfrac{R}{\ell_{\mathsf{dS}}}  + \beta_{\rm s}(T)\left(1-\cfrac{R^2}{\ell_{\mathsf{dS}}^2}\right)^{1/2} \right)^2 \dd T^2  ~,
\ee
with $R< \ell_{\mathsf{dS}} $, and $\alpha_{\rm s}(T)$ and $\beta_{\rm s}(T)$ are arbitrary functions. This spacetime is well adapted to the static patch and locally dS$_2$. 

\section{Kicking the ultracold black hole} \label{sec:ultracold}

We finally turn to the most novel instance of our extremal cases: the ultracold black hole.  Recall that this solution is obtained when all the three horizons coincide
\be \label{eq:equal-horizons}
r_-=r_+=r_c\equiv r_{\mathsf{uc}}~,
\ee
and it corresponds to the point in phase space where
\be 
\begin{aligned}
r_{\mathsf{uc}}= \frac{\ell_4}{\sqrt{6}}~, \qquad
Q_{\mathsf{uc}}^2 = \frac{\ell_4^2}{12}~,  \qquad 
M_{\mathsf{uc}}^2= \frac{2\ell_4^2}{27}~.
\label{UCext}
\end{aligned}
\ee
This is the first peculiarity of this solution: while in the previous cases extremality can be obtained by choosing different values of $r_0$ or $r_{\mathsf{n}}$ (thereby different values of $M$ and $Q$ according to \eqref{ZMRomans} and \eqref{ZMN1}), the ultracold case is more constrained: the extremal solution corresponds to a single point. The point is located at the tip of the Shark Fin in Fig.\,\ref{SharkFin}: one sees immediately then that moving horizontally (namely, raising the mass, keeping the charge fixed) corresponds to going out of the shaded area, and encountering a naked singularity.  
  \begin{wrapfigure}{r}{0.38\textwidth}
	\centering
	\includegraphics[width=\linewidth]{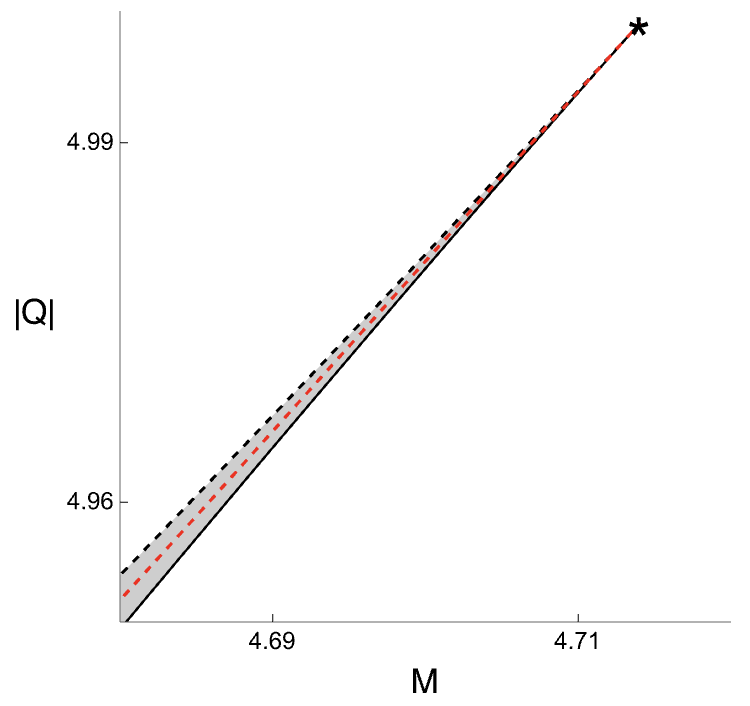}
	\caption{Close-up of Fig.\,\ref{SharkFin}, near to the ultracold black hole.} 
	\label{sharkTip}
	\end{wrapfigure}
 
Our strategy for ``heating up'' the ultracold solution will then be different: we should work in an ensemble that allows charge and mass to vary, while keeping us inside the Shark Fin. In other words, we should allow a near-extremal deformation that moves the solution downwards along the red line displayed in Fig.\,\ref{sharkTip}. 

The next subsections are devoted to explaining how to achieve this, and what are the consequences of this procedure. That is, we will kick the ultracold black hole away from extremality.  We will first investigate the consequences at the level of black hole thermodynamics and the near-horizon geometry. We will then carry out the holographic analysis from the two-dimensional perspective, by analyzing the  perturbations around  Mink$_2 \times S^2$; we will also show how they match with the black hole response.

\subsection{Near-extremality: thermodynamics and geometry}\label{sec:near-uc}

\paragraph{Thermodynamics.} As familiar by now, the deviation away from extremality is given by introducing $\lambda$ to split the coincident horizons in \eqref{eq:equal-horizons} by a small amount. In the context of a thermodynamic analysis, we will first investigate how $Q$ and $M$ respond to a deviation away from the cold solution. We start by first sending 
\be 
r_-= r_{\mathsf{uc}}-w_1\lambda+O(\lambda^2)~,\qquad r_+= r_{\mathsf{uc}} - w_2 \lambda+O(\lambda^2)~,
\label{rpmw1w2}
\ee
where $w_2<w_1$ are constant coefficients and finite as $\lambda\to 0$.\footnote{At this stage we only ask that $w_2<w_1$, so that $r_-<r_+$ at leading order in $\lambda$.} We will also be holding fixed $\ell_4$, and hence from \eqref{constraints} we can quantify  the response of $Q$ and $M$; we find, 
\be
\begin{aligned}
Q=Q_{\mathsf{uc}}-\frac{2 }{\sqrt{3} \, \ell_4}(w_1^2+w_1w_2+w_2^2)\lambda^2 +O(\lambda^3)~,\\
\label{z}
M =  M_\mathsf{uc}-\frac{\sqrt{2}}{\sqrt{3}\, \ell_4} (w_1^2+w_1 w_2+w_2^2) \, \lambda ^2+O(\lambda ^3)~.   
\end{aligned}
\ee
This clearly illustrates the basic differences relative to cold and Nariai black holes. First, the leading response is order $\lambda^2$, rather than $\lambda$ for arbitrary $w_{1,2}$. Second, there is no real values of $w_1$ and $w_2$ that will hold $Q$ fixed at leading order. This is compatible with the intuition gathered from Fig.\,\ref{sharkTip}; for any value of $w_{1,2}$ and small $\lambda$, the deviations \eqref{z} are within the shark fin. 

Next, it is instructive to quantify how the temperature at each horizon responds to these deviations. For this, it is first useful to record that 
\be 
r_c= r_{\mathsf{uc}}+(w_1+w_2)\lambda+O(\lambda^2)~.
\label{rcw1w24}
\ee
With this we assure that \eqref{rpmw1w2} and \eqref{rcw1w24} leave $\ell_4$ fixed at leading order in $\lambda$. The responses of the cosmological and outer horizons to the deviations in  \eqref{rpmw1w2} and \eqref{rcw1w24} give
\be 
\begin{aligned}
 T_c&=\frac{\sqrt{6}}{\pi\, \ell_4^3} \left(2w_2+w_1\right)\left(2w_1+w_2\right)  \lambda^2+O(\lambda^3)~,\\
 T_+&= \frac{\sqrt{6}}{\pi \ell_4^3}\left(2w_1+w_2\right)\left(w_1-w_2\right)\lambda^2+O(\lambda^3)~.
\label{TUC}
\end{aligned}
\ee
Again this is very different from our previous situations: for the cold and Nariai black holes the response in the temperature was $T_\mathsf{h}\sim O(\lambda)$, while here we obtain $T_\mathsf{h}\sim O(\lambda^2)$. 

Actually, the quantities that respond at leading order in this scenario are the electric potential and the entropy. In particular we find 
\be 
\Phi_{\mathsf{c}}= \frac{Q}{r_c}= \frac{1}{\sqrt{2}} - \frac{\sqrt3}{\ell_4}(w_1+w_2) \lambda +O(\lambda^2)~,
\label{phii}
\ee
and the area law at $r_c$ is 
\be 
S_c=\frac{\pi \, \ell_4^2 }{6}+\sqrt{\frac{2}{3}} \pi \, \ell_4 (w_1 +w_2)\lambda  +O(\lambda^2)\,.
\label{suc}
\ee
From these expressions it is natural to advocate that 
 the change in entropy at order $\lambda$ is driven by a change of chemical potential rather than to a change in temperature (which is subleading). This is reminiscent of other studies of two-dimensional gravity theories in flat spacetime \cite{Afshar:2019axx}, where an infinite value for the specific heat
\begin{equation} \label{spec_heat}
    C_s^{-1} = \frac{1}{T} \left(\frac{dT}{dS} \right) \bigg|_{Q={\rm const}}\,
\end{equation}
was found, since the change in temperature is independent on the change in entropy. The subsequent portions are devoted to showing how to retrieve this feature via an analysis of the IR background for the ultracold case.

\paragraph{Near-horizon geometry.}

 After displacing the location of the horizons following \eqref{rpmw1w2} and \eqref{rcw1w24}, we will now construct the near-horizon geometry.  To keep expressions simple and succinct, and without loss of generality, we will make a specific choice of $w_{1,2}$: setting $w_1 =\epsilon$ and $w_2=0$, we have
\be
r_- = r_{\mathsf{uc}}  -  \epsilon  \, \lambda~,
\qquad r_+ = r_{\mathsf{uc}}~,   \qquad 
r_ c =  r_{\mathsf{uc}} + \epsilon \,\lambda ~.
\ee
This is different than the deviation used in \eqref{bro}, since in that instance the solution was still extremal. The coordinate transformation we will use is
\be\label{eq:dec-ucold}
\begin{aligned}
r&= r_{\mathsf{uc}} -{R_0}\lambda + \, \lambda ^{3/2}   \sqrt{\frac{2R_0^3 }{3 r_\mathsf{uc}^{3}}} \, R ~, \\ 
t &=  \sqrt{\frac{3 r_\mathsf{uc}^{3}}{2 R_0^3} } \frac{T}{\lambda ^{3/2} } ~,
\end{aligned}
\ee
where $R_0$ is an arbitrary constant. With this choice the resulting near-horizon background is
\be \label{eq:ext-ucold-2}
ds^2=-\cfrac{R_0^2-\epsilon^2}{ R_0^2}\, \dd T^2+ \cfrac{R_0^2}{R_0^2-\epsilon^2}\, \dd R^2+r_\mathsf{uc}^2 \,\left(\dd\theta^2+\sin^2\theta\dd\phi^2\right)~,
\ee
and
\be
F= \pm \frac{\sqrt{3}}{\ell_4} \dd T \wedge \dd R~.
\ee
For $\epsilon =0$ we recover \eqref{eq:ext-ucold-1} and \eqref{eq:ext-ucoldF-1}, as expected. However notice that the presence of $\epsilon$ is trivial: it can be completely reabsorbed by a constant rescaling of $T$ and $R$. In this context it is clear that a deviation from extremality is {\it not} heating up Mink$_2$.\footnote{Moreover, have we obtained in \eqref{eq:ext-ucold-2} a finite temperature solution in the IR, it would imply that in UV $T_\mathsf{h}\sim O(\lambda^{3/2})$ due to \eqref{eq:dec-ucold}. However, we know that in the UV $T_\mathsf{h}\sim O(\lambda^{2})$ as discussed in \eqref{TUC}.}

For the comparison with the subsequent holographic analysis it is useful to transform the coordinates to
\be\label{eq:tuRr}
T = u +R~, \qquad R = \hat{r}~,
\ee
which brings the metric $\overline{g}_{ab}$ in the Eddington-Finkelstein form
\be\label{eq:yy1}
ds^2 = - \dd u^2 -2 \dd u \dd\hat{r}~.
\ee
Note that in \eqref{eq:tuRr} we have rescaled \eqref{eq:ext-ucold-2} to have the usual normalization of Mink$_2$. The first correction in $\lambda$ are also simple to cast. With  our choice of coordinates we have
	\begin{equation}
	\begin{aligned} \label{expans_UC}
	ds^2&= \left(\bar{g}_{ab} +\sqrt{\lambda} \, \tilde{g}_{ab} + \lambda\,  h_{ab}\right)  \dd x^a \dd x^b+ \left(\Phi_0^2 + 2\lambda \Phi_0 \, Y \right)\left(\dd\theta^2+\sin^2\theta\dd\phi^2 \right)+\cdots ~,
 \\
 A &= \bar A_{a} \dd x^a + \lambda\, {\cal A}_a \dd x^a + \cdots~.
	\end{aligned}
	\end{equation}
One difference, relative to cold and Nariai, is that we get a correction that grows like $\lambda^{1/2}$. This is simply an artifact of our decoupling limit; it is straightforward to show that $\tilde{g}_{ab}$ is pure gauge. For the linear corrections we find
\be\label{eq:yy2}
\Phi_0= r_{\mathsf{uc}}~,\qquad Y(x)= -R_0~.
\ee
In contrast to cold and Nariai, here we have a constant profile for the deformation.

\subsection{Holographic analysis}
In this section we will account for the unusual thermodynamic behaviour of the ultracold backgrounds by doing the holographic analysis of Mink$_2$. This discussion follows the treatment of \cite{Afshar:2019axx,Godet:2021cdl,Grumiller:2021cwg}, which considered theories of two-dimensional dilaton-gravity that admit a flat space vacuum. The benefit in our case is that we have a black hole embedding for Mink$_2$, and hence we can systematically compare and contrast the two-dimensional results with a four-dimensional realization.

Our starting point is to consider as a background solution the two-dimensional Minkowski space metric. This is the solution described in Sec.\,\ref{sec:JTreduction}, where 
\be\label{eq:bbuc}
\Phi_0^2 = \frac{\ell_4^2}{6}~, \qquad Q^2= Q^2_{\mathsf{uc}}=\frac{\Phi_0^2}{2}\,.
\ee
We will cast a locally Mink$_2$ space in Eddington-Finkelstein coordinates
\be \label{edd-fink}
ds^2 = - 2 \left( \mathcal{P}(u) \hat{r} + \mathcal{T} (u) \right) \dd u^2-2 \dd u \dd \hat{r}~,
\ee
and the field strength is given by the volume form of this space and normalized according to \eqref{eq:F1}. For most of the discussion we leave the functions  $\mathcal{P}$ and $\mathcal{T}$ general and dependent on $u$. However for the near-horizon geometry displayed in the previous section, they are both constant: we will show  explicit solutions specializing to constant values of $\mathcal{P}$ and $\mathcal{T}$ which we denote by ${\cal P}_0$ and ${\cal T}_0$.

In the following we will discuss certain dynamical aspects that arise when the background solution is deformed away from its fixed point. We will first solve for linear perturbations around this background, and then quantify their imprint on the renormalized action.  An important aspect will be to  contrast basic properties here against those for  AdS$_2$ in Sec.\,\ref{sec:hol-cold}: the interplay between $Y(x)$ and the background  solution will not play a central role for Mink$_2$.

\subsubsection{Perturbations around \texorpdfstring{Mink$_2$}{Mink2} } \label{pert_UC}

The aim here is simple: to solve the linear equations \eqref{massaged1} and \eqref{massaged2} when the background is given  by \eqref{eq:bbuc}-\eqref{edd-fink}. Starting from \eqref{massaged1}, which determines the dynamics of the dilaton,  we find 
\be
\begin{aligned}\label{UCY}
     \partial_{\hat{r}}^2   Y (x) & = 0  \\
 \partial_u \partial_{\hat{r}} Y -  \mathcal{P}(u) \partial_{\hat{r}} Y    & =  \frac{ \delta Q}{\sqrt{8}Q^2}  \\
 \left( {\hat{r}} \mathcal{P}\,'(u) + \mathcal{T}\,'(u) \right) \partial_{\hat{r}} Y - \mathcal{P}(u) \partial_u Y +2 ({\hat{r}} \mathcal{P}(u) + \mathcal{T}(u)) \, \partial_u \partial_{\hat{r}} Y - \partial_u^2 Y & =  0~. 
\end{aligned}
\ee
Notice that we have used \eqref{eq:bbuc} to simplify these expressions. For metric perturbations, we will make a choice of gauge where $h_{\hat{r}u}=h_{\hat{r}\hat{r}}=0$; this greatly simplifies \eqref{massaged2}, leaving us with 
\be
 \partial_{\hat{r}}^2 h_{uu} - \frac{\sqrt{2}}{Q^3} \, Y(x) +\frac{3 \,  \delta Q}{2 Q^3}   =  0~. \label{UCmetricEQ}
\ee
The solution to these equations are simple to decode. From the first equation in \eqref{UCY} we read off the radial profile of $Y(x)$,  which is
\be
Y = a(u) \hat{r} + b(u)~.
\ee 
The functions $a(u)$ and $b(u)$ are determined by the two last equations in \eqref{UCY}, and these lead to
\be
\begin{aligned}\label{eq:ab1}
a'(u)- a(u) \mathcal{P}(u)- \frac{\delta Q}{{2 \sqrt2} Q^2}  =0 ~,\\
{\cal T}'(u) a(u)-{\cal P}(u)b'(u)+2{\cal T}(u) a'(u)-b''(u)=0~.
\end{aligned}
\ee
 The inhomogeneous solution to \eqref{UCmetricEQ} is given by
\be
h_{uu} =r^3 \frac{ a(u)}{3 \sqrt2 \, Q^3}  + r^2\frac{ \, b(u)}{\sqrt2 \, Q^3} -r^2\frac{3 \, \delta Q   }{4\, Q^3}~.
\ee

It is useful to explicitly record the solutions to the equations displayed above when the Mink$_2$ is independent of $u$. There are two distinct  branches of solutions:
\begin{description}[leftmargin=0.2cm]
    \item[Branch I.] This branch corresponds to solutions the Mink$_2$ backgrounds with   ${\cal P}(u)=0$, and ${\cal T}(u)=1$. The solutions to \eqref{eq:ab1} are
\be\label{eq:ysol11}
a(u)= \frac{\delta Q}{{2 \sqrt2} Q{^2}} \, u +a_0~, \qquad b(u)=\frac{\delta Q}{{2 \sqrt2} Q{^2}}\, u^2 + b_1 u + b_0~,  
\ee
where $a_0$ and $b_{1,0}$ are arbitrary constants. In comparison to \eqref{eq:yy1} and \eqref{eq:yy2}, relevant for the near-horizon geometry of the ultra-cold black hole, we have $\delta Q=a_0=b_1=0$. Then, $b_0=R_0$ is the only non-trivial component of the dilaton.
\item[Branch II.] In this case the metric functions are ${\cal P}(u)={\cal P}_0$ and ${\cal T}(u)={\cal T}_0$, but  with the important caveat that ${\cal P}_0$ non-zero. This gives
\be
\begin{aligned} \label{YsolTOTUC}
a(u)&= -\frac{1}{{2 \sqrt2} {\cal P}_0}\frac{\delta Q}{ Q{^2}}  +a_1 e^{{\cal P}_0 u}~,\\
   b(u)&=  a_1 \frac{{\cal T}_0}{{\cal P}_0} e^{{\cal P}_0 u}+ b_2\, e^{-{\cal P}_0 u} +b_0~. 
   \end{aligned}
\ee
Here $a_{1}$, $b_{2}$ and $b_0$ are arbitrary constants. 
 The solution has the same form as that found in \cite{Godet:2021cdl} in models of $\widehat{\rm CGHS}$ gravity. However, the near-horizon geometry \eqref{eq:yy1} and \eqref{eq:yy2} does not belong to this branch.
\end{description}

It is worth highlighting features of branch I and II solutions. 
\begin{enumerate}
    \item A static, $u$-independent, solution of branch I requires   $\delta Q=0$ and $b_1=0$ in \eqref{eq:ysol11}.  Moreover, equations \eqref{eq:ab1} become background independent. 
    This is very different from the analogous condition for AdS$_2$ in Sec.\,\ref{sec:hol-cold}, where fixing $Q$ did not affect significantly the JT field.
    \item For branch II, static solutions can have $\delta Q$ non-zero. However, if  we set $\delta Q=0$, the profile of $Y(x)$ is again independent of the background metric. This is also different from the AdS$_2$ counterpart.  
    \item  Following \cite{Godet:2021cdl}, if we impose the boundary condition
 \be
 Y(x) \xrightarrow[\hat{r} \to \infty]{} \Phi_r \hat{r}~,
\ee 
where $\Phi_r$ is fixed, arbitrary charge variations are not allowed. We would have to require 
\be\label{eq:bc-godet}
\delta Q = -{2 \sqrt2} Q^2  {\cal P}_0 \Phi_r ~.
\ee
This will have an imprint in the thermodynamics discussed below. 
\end{enumerate}
These features can be taken as an indication that there is a strange interplay between the deformation $Y(x)$ and heating up Mink$_2$.


\subsubsection{Thermodynamics around \texorpdfstring{Mink$_2$}{Mink2}}

With the solution for the perturbations at hand, we can compute thermodynamic quantities associated to  the two-dimensional black hole, with the aim to connect with the near-extremal thermodynamics of the ultracold black hole in Sec.\,\ref{sec:near-uc}. For a static two-dimensional solution we have to set to zero all terms which are $u$ dependent. For the IR background, this means that we have
\be \label{edd-fink1}
ds^2 = - 2 \left( \mathcal{P}_0 \hat{r} + \mathcal{T}_0 \right) \dd u^2-2 \dd u \dd \hat{r}~.
\ee
We will interpret this as a ``black hole'' solution, whose horizon is at $\hat{r}_h=-{\cal T}_0/{\cal P}_0$. The associated Hawking temperature is \cite{Afshar:2019axx}
\begin{equation}
    T_{\rm 2D} = \frac{{\cal P}_0}{2 \pi}\,,
\end{equation}
which is defined as the surface gravity of the Killing vector $k=\partial_u$. Therefore branch I solutions are always cold $(T_{\rm 2D}=0)$, and branch II solutions can be warm ($T_{\rm 2D}\neq0$). 

A static configuration for the dilaton corresponds to 
branch II, ${\cal P}_0\neq0$, means setting $a_1=0$ and $b_2=0$ in \eqref{YsolTOTUC};  this gives
\be\label{eq:Ystatic-uc}
\begin{aligned}
   {\rm \bf Branch~~ I:}\qquad &Y(x)=  b_0~,\\
   {\rm \bf Branch~~ II:}\qquad &Y(x)= -\frac{1}{{2 \sqrt2}{\cal P}_0}\frac{\delta Q}{Q{^2}} \hat{r} + b_0~,\\    
\end{aligned}
\ee
which follows from \eqref{eq:ysol11} and \eqref{YsolTOTUC}.
Notice that the boundary condition \eqref{eq:bc-godet} now can be interpreted as $\delta Q\sim T_{\rm 2d}$. For both branches, the entropy is dictated by the value of the dilaton $Y$ evaluated at the horizon, for which we obtain
\be
\begin{aligned}\label{eq:entropy-mink2}
S_{2D} &= \pi \Phi(x)^2_{\rm horizon}\\
&= \pi \Phi_0^2 + 2\pi \Phi_0   \lambda Y(x)_{\rm horizon}+\cdots\\
     &=  \pi \Phi_0^2 + 2\pi \Phi_0  b_0  \lambda -   \frac{{\cal T}_0}{T_{\rm 2D}} \Phi_0 \Phi_r \lambda  + \cdots
\end{aligned}
\ee
where in the last line we replaced \eqref{eq:bc-godet}. The term controlled by $b_0$ is consistent with the behaviour of the 2d perturbations described by models of dilaton gravity in flat spacetime such as $\widehat{\rm CGHS}$ \cite{Afshar:2019axx,Grumiller:2021cwg}. Indeed, when compared with the ultracold black hole entropy \eqref{suc}, we see no dependence of the entropy variation on the change in temperature, hence $C_s$, as defined in \eqref{spec_heat}, is infinite.  

The last term in \eqref{eq:entropy-mink2} is clearly strange and should be treated carefully. First, notice that there is an important order of limits: if $T_{\rm 2D}=0$, i.e., ${\cal P}_0=0$, we are in branch I and hence only $b_0$ gives a contribution to the entropy. For this reason we do not see this contribution in \eqref{suc}. There are at least three ways to ``normalize'' this divergent behaviour for ${\cal P}_0\neq0$: one could set ${\cal T}_0=0$ as done in \cite{Godet:2021cdl}, set $\delta Q=0$, or modify the boundary condition \eqref{eq:bc-godet} such that $\delta Q\sim T_{\rm 2D}^2$.\footnote{Formally, it would be interesting to modify \eqref{eq:bc-godet} and study more carefully its repercussions. Unfortunately we don't see an indication that a modified boundary condition for $\delta Q$ is appropriate for the ultracold black hole, so we leave this for future work. }

To complete the picture we perform holographic renormalization to compute the on-shell action. Replacing the solution we found in Sec.\,\ref{pert_UC} into the 2D action \eqref{eq:2daction} gives a divergent result for the on-shell action. We regulate the integral by introducing an extremum of integration $\hat r_0$ taken to be large but finite (UV cutoff), while the other extremum of integration is the black hole horizon located at $\hat r=\hat r_h$. To remove the divergences we add the following counterterms:
\be
I_{\rm on-shell-uc} = I_{\rm 2D}+I_{\rm N} + I_{\rm GH} + I_{\rm MM}\,.
\ee
The first term is the action \eqref{eq:2daction}. The  subsequent  terms are  
\begin{equation}
   I_{\rm N}  = {{-\frac{1}{4}}} \int \dd u \, \sqrt{-\gamma} \, \Phi  (n^{a} \partial_{a} \Phi) ~,  \qquad   I_{\rm GH}  = {\frac{1}{2}} \int \dd u\,\sqrt{-\gamma} \, \Phi^2  \, K ~,
\end{equation}
where $\gamma_{ab}$ is the boundary metric and $n^a$ is the unit vector normal to the boundary. $I_{\rm N}$ is a standard counterterm for models of dilaton- gravity in flat 2d space (see for instance \cite{Godet:2021cdl,Kar:2022sdc,Kar:2022vqy}) and $I_{\rm GH}$ is the standard Gibbons-Hawking-York term, which ensures Dirichlet boundary conditions for the metric. As usual in flat space, we have to supplement this action by the Mann-Marolf boundary term \cite{Mann:2005yr},
\be
I_{\rm MM} = -\frac{1}{{2}} \int \dd u\,  \sqrt{-\gamma_{\rm ref}}\, \Phi^2 \, \hat{K}_{\rm ref}~, 
\ee
with representative
\be
ds^2_{\rm ref} = -2\dd u\dd r - 2(\mathcal{P}(u) r_0 +\mathcal{T}(u))\dd u^2{+ O(\lambda)}~,
\ee
where $\hat r_0$ is the radial (UV) cutoff. Adding this term effectively amounts to performing a  background subtraction and in this way the action is free of divergences. 

The final finite expression for the renormalized action boils down to
\be\label{onsh_intermediate}
I_{\rm on-shell-uc} = \lambda 
\Phi_0 \int \dd u  \left[ \left(b(u) \mathcal{P}(u) - a(u) \mathcal{T}(u)  \right) + \frac{Q \mu(u)}{\Phi_0^3} \right] +I_{\rm global}\,.
\ee
 A similar form for the on-shell action (which does not include the chemical potential term) was found in \cite{Godet:2021cdl}. 

With \eqref{onsh_intermediate} we can now extract the entropy of the two-dimensional black hole. We will be interested in the case where $\delta Q=0$, and we will also take ${\cal P}_0\neq0$: these choices will facilitate comparison with the ultracold black hole and we will be able to take ${\cal P}_0\to0$ smoothly. Evaluating \eqref{onsh_intermediate} on \eqref{edd-fink1}-\eqref{eq:Ystatic-uc},    the Euclidean action then gives,
\be\label{eq:on-shell-uc11}
I_{\rm on-shell-uc} = -2\pi \Phi_0    b_0 \lambda+I_{\rm global}\,,
\ee
where we set $\delta Q=0$. 
This expression clearly reflects that the temperature does not affect the on-shell action, and hence the entropy, as we have been expecting. Using the standard thermodynamic relation
\be
S = \beta \left( \frac{\partial I}{ \partial \beta} \right) -I ~,
\ee
we find that $S_{\rm 2D}=-I_{\rm on-shell-uc}$ is in agreement with  \eqref{eq:entropy-mink2} for fixed charge, and agrees with the response of the ultracold black hole. We emphasise that this is very different from the outcome in Sec.\,\ref{sec:hol-cold}, where temperature is the leading effect in the deformation.

\section{Discussion}\label{sec:conclusion}

We have established the basic ingredients to describe the thermodynamic properties of RNdS$_4$ black holes near to extremality. We mainly focused on the linear response around the three extremal limits in consideration: cold, Nariai and ultracold. We provided two different perspectives on this: a four-dimensional analysis, which is explicitly drawn from the 4D de Sitter black hole thermodynamics, and a two-dimensional effective theory, which efficiently captures the dynamics of the fields obtained in the Kaluza-Klein reduction. 

Below we highlight some of our results combined with some possible future directions. 

\paragraph{Non-linear effects.} Although we used the two-dimensional theory \eqref{eq:2daction} only at the linear level, it is a consistent truncation at the non-linear level. It would be interesting to investigate non-linear aspects of the extremal responses, where the effects of the dS$_4$ surrounding become important, for instance deducing the shape of the fin in Fig.\,\ref{SharkFin} from the two-dimensional perspective. More concretely, one would like to discard branch II in \eqref{YsolTOTUC} as a possibility once non-linear effects are incorporated; recall that branch II is incompatible with the thermodynamics of RNdS$_4$. 

\paragraph{dS$_2$ analysis and the Nariai limit.} Our perspective of the Nariai limit was to view it as a zero temperature configuration where dS$_2$ is cold. This choice does not agree with the proposal of \cite{Bousso:1996au}, where time is rescaled such that the temperature remains finite in the Nariai limit. For our purposes, this rescaling introduces tension in the thermodynamic response near-extremality: although the temperature is finite, the electric potential diverges. We believe our setup is more robust, although it would be interesting to investigate these different perspectives and scrutinise them from the two-dimensional theory. Recent work which explore the Nariai limit, and follows the prescription of  \cite{Bousso:1996au}, are \cite{Svesko:2022txo,Morvan:2022aon}.  

\paragraph{Exploring de Sitter Black Holes.} Our results could be extended and applied in several different directions. We expect this to extend to higher-dimensional RNdS$_d$ background rather easily. A more rich direction is to extend our analysis to rotating black holes in de Sitter. Kerr-dS$_4$ also has a Nariai and ultracold limit that would be interesting to revisit in the near extremal limit along the lines of \cite{Anninos:2010gh}, and to connect with the Kerr/CFT correspondence proposed in \cite{Anninos:2009yc}.

\paragraph{Other methods to explore perturbations.}  It would be interesting to complement our two-dimensional perspective with other methods used to study perturbations of the RNdS$_4$ solution. For example, to connect to the more general studies in \cite{Dias:2018etb} that study strong cosmic censorship, and to find the 2D realization of the Festina Lente bound \cite{Montero:2019ekk}. It would also be interesting to explore the stability properties of the ultracold solution along the lines of \cite{Anninos:2022hqo,Horowitz:2022mly}.

\section*{Acknowledgements}

We thank Dio Anninos, Tarek Anous, Stephane Detournay, Albert Law, Andrew Svesko,  and Evita Verheijden for interesting discussions, and collaborations on related topics.
The work of AC and CT was supported by the Delta ITP consortium, a program of the Netherlands Organisation for Scientific Research (NWO) that is funded by the Dutch Ministry of Education, Culture and Science (OCW). The work of AC has been partially supported by STFC consolidated grant ST/T000694/1. The work of CT is supported by the Marie Sk\l odowska-Curie Global Fellowship (ERC Horizon 2020 Program) SPINBHMICRO-101024314. FM acknowledges financial support from the European Research Council (grant BHHQG-101040024) funded by the
European Union. Views and opinions expressed are however those of the author(s) only and do not
necessarily reflect those of the European Union or the European Research Council. Neither the European Union nor the granting authority can be held responsible for them.

\providecommand{\href}[2]{#2}\begingroup\raggedright\endgroup

\end{document}